\PassOptionsToPackage{square,comma,numbers,sort&compress,super}{natbib}  
\documentclass{article}

\pdfoutput=1 
\usepackage{arxiv}
 
\usepackage[utf8]{inputenc} % allow utf-8 input
\usepackage[T1]{fontenc}    % use 8-bit T1 fonts
\usepackage{hyperref}       % hyperlinks
\usepackage{url}            % simple URL typesetting
\usepackage{booktabs}       % professional-quality tables
\usepackage{amsfonts}       % blackboard math symbols
\usepackage{nicefrac}       % compact symbols for 1/2, etc.
\usepackage{microtype}      % microtypography

\usepackage{hyperref}
\usepackage{times}
\usepackage{fancyhdr}
\usepackage{makecell}
\usepackage{bm}
\usepackage{float}
\usepackage{filecontents}
\usepackage[linesnumbered,ruled]{algorithm2e}
\usepackage[noend]{algpseudocode} 
\usepackage[dvipsnames]{xcolor}
\usepackage{amsmath}
\usepackage{amssymb}
\usepackage{subfigure}
\usepackage{multirow}
\usepackage{enumitem}  
\usepackage{comment}
\usepackage[pdftex]{graphicx}  

\hypersetup{
	colorlinks=false,
	pdfborder={0 0 0},
}

\title{Block-Based Spectral Processing of Static and Dynamic 3D Meshes using Orthogonal Iterations}

\author{
  Gerasimos ~Arvanitis\\
  Department of Electrical and Computer Engineering\\
  University of Patras\\
  Greece \\
  \texttt{arvanitis@ece.upatras.gr} \\
   \And
  Aris S.~Lalos \\
  Industrial Systems Institute\\
  ATHENA Research and Innovation Center\\
  Platani-Patras, 26504, Greece \\
  \texttt{lalos@isi.gr} \\
  \And
  Konstantinos ~Moustakas\\
  Department of Electrical and Computer Engineering\\
  University of Patras\\
  Greece \\
  \texttt{moustakas@ece.upatras.gr} \\
}

\begin{document}
\maketitle

\begin{abstract}
Spectral methods are widely used in geometry processing of 3D models. They rely on the projection of the mesh geometry on the basis defined by the eigenvectors of the graph Laplacian operator, becoming computationally prohibitive as the density of the models increases. In this paper, we propose a novel approach for supporting fast and efficient spectral processing of dense 3D meshes, ideally suited for real-time compression and denoising scenarios. To achieve that, we apply the problem of tracking graph Laplacian eigenspaces via orthogonal iterations, exploiting potential spectral coherence between adjacent parts. To avoid perceptual distortions when a fixed number of eigenvectors is used for all the individual parts, we propose a flexible solution that automatically identifies the optimal subspace size for satisfying a given reconstruction quality constraint. Extensive simulations carried out with different 3D meshes in compression and denoising setups, showed that the proposed schemes are very fast alternatives of SVD based spectral processing while achieving at the same time similar or even better reconstruction quality. More importantly, the proposed approach can be employed by several other state-of-the-art denoising methods as a preprocessing step, optimizing both their reconstruction quality and their computational complexity.
\end{abstract}

% keywords can be removed

\section{Introduction}
\label{intro}

In recent years, there has been increasing interest from researchers, system designers, and application developers on acquiring, processing, transmitting and storing 3D models, facilitating several real-time applications, e.g., mobile cloud gaming \cite{Cai2013} and 3D tele-immersion \cite{Alexiadis2013},\cite{Mekuria2014}. These models usually come as very large and noisy meshes that stand in need of solutions for a diversity of problems including mesh compression, smoothing, symmetry detection, watermarking, surface reconstruction, and re-meshing \cite{zhang2010spectral}. Spectral methods have been developed with the intention of solving such problems by manipulating the eigenvalues, eigenvectors, eigenspace projections, or a combination of these, derived from the graph Laplacian operator. The processing and memory requirements of these methods are strongly dependent on the number of vertices of the 3D model, and therefore become prohibitive as the vertex density increases, especially in cases where the models are too large and need to be scanned in parts, generating a sequence of 3D surfaces that arrive sequentially in time. To address this issue, the raw geometry data could be divided and processed in blocks that represent the different parts of a mesh (submeshes), as suggested in \cite{Lalos_tmm},\cite{lalos2015sparse}.

The application of direct singular value decomposition (SVD) on the graph Laplacian of each submesh, requires $\mathcal{O}\left(n_d^3\right)$ operations, where $n_d$ is the number of vertices in a submesh. This excessively high computational complexity needed by SVD motivated us to seek for an efficient subspace tracking implementation that processes the raw geometry data in blocks and readjust only a small number of spectral coefficients of a submesh based on the corresponding spectral values of a previous submesh. The proposed approach is based on a numerical analysis method known as orthogonal iterations (OI) \cite{Zhangthesis}, which is capable of estimating iteratively the subspaces of interest. The speed-up is attributed to the fact that the proposed approach requires $\mathcal{O}\left(n_dc^2\right)$ floating point operations where $c$ is the number of spectral components utilized and $c<<n_d$. Additionally, we developed a dynamic OI approach that estimates automatically the ideal $c$ for a predefined reconstruction quality. Extensive simulations carried out with different 3D meshes in a compression and de-noising setup, proved that the proposed framework is a very fast alternative of the SVD based graph Laplacian processing methods, without introducing noticeable reconstruction errors. 

The rest of the article is organized as follows. Section 2 provides a review of prior art on spectral methods and their applications in a diversity of problems. Basic definitions related to graph spectral processing of 3D meshes are provided in Section 3. Section 4 presents the proposed fast spectral processing approach that is based on OI. Section 5 provides a flexible solution that automatically identifies the optimal subspace size $c$ that satisfies a specific reconstruction quality criterion. In Section 6 we investigate the spatial coherence between submeshes of the same mesh. We also study the impact of the submesh size to the reconstruction quality and the computational complexity of the proposed approach. Section 7 presents a compression and a denoising case study, where the proposed method can be effectively adopted. In Section 8, the performance of the proposed system is evaluated, by taking into account different CAD and scanned 3D models. The article is finally wrapped up with a few open research directions in Section 9.

\section{Related Works}
\label{relatedworks}
Spectral methods have been used in many different computer science fields ranging from signal processing, graph theory, computer vision and machine learning. Spectral mesh processing have been inspired by all the relevant developments in the aforementioned fields. Several surveys that cover basic definitions and applications of the graph spectral methods have been introduced by Gotsman \cite{gotsman2003graph}, Levy \cite{levy2006laplace}, Sorkine \cite{Sorkine2005} and more recently by Zhang et al. \cite{zhang2010spectral}.  All these surveys classify the spectral methods according to several criteria related to the employed operators, the application domains and the dimensionality of the spectral embeddings used.

Graph spectral processing of 3D meshes rely on the singular/eigenvectors and/or eigenspace projections derived from appropriately defined mesh operators, while it has been applied in several tasks, such as, implicit mesh fairing \cite{kim2005geofilter}, geometry compression \cite{Sorkine2005, Karni2000} and mesh watermarking \cite{ohbuchi2001watermarking}. Taubin \cite{Taubin-1995} first treated the mesh vertex coordinates as a 3D signal and introduced the use of graph Laplacian operators for discrete geometry processing. This analysis was motivated by the similarities between the spectral analysis with respect to mesh Laplacian and the classical Fourier analysis. A summary of the mesh filtering approaches that can be efficiently carried out in the spatial domain using convolution approaches is given by Taubin in \cite{taubin2000}. Despite their applicability in a wide range of applications such as mesh denoising, geometry compression and watermarking, they require explicit eigenvector computations making them  prohibitive for real time scenarios. Additionally, there are a lot of applications in literature in which large-scale 3D models are scanned in parts \cite{7831815}, \cite{7949103}, \cite{7027835} providing a sequence of 3D surfaces that need to be processed fast and sequentially in time. Our method has been designed in order to be ideally suited particularly in these cases, providing accurate results while the whole process takes part in real-time. 

Computing the truncated singular value decomposition, can be extremely memory-demanding and time-consuming. To overcome this limitations, \emph{subspace tracking} algorithms have been proposed as fast alternatives relying on the execution of iterative schemes for evaluating the desired eigenvectors per incoming block of floating point data corresponding in our case, to different surface patches~\cite{Comon1990}. The most widely adopted subspace tracking method is Orthogonal Iterations (OI), due to the fact that results in very fast solutions when the initial input subspace is close to the subspace of interest, as well as the size of the subspace remains at small levels~\cite{Saad2016}. The fact that both matrix multiplications and QR factorizations have been highly optimized for maximum efficiency on modern serial and parallel architectures, makes the OI approach more attractive for real time applications. 

This work is an extended version of the research presented in \cite{grapp18}. In this version, we provide more details about the ideal mesh segmentation (e.g., number of submeshes, size of overlapped submehses) and the submeshes properties (e.g., spatial coherence between submeshes of the same mesh). Additionally, we extend the application scenarios presenting a block-based spectral denoising approach for 3D dynamic meshes.

\section{Spectral Processing of 3D Meshes}
\label{sec:GFT3dD}

In this work we focus on polygon models whose surface is represented using triangles. Let us assume that each triangle mesh $\mathcal{M}$ with $n$ vertices can be represented by two different sets $\mathcal{M}=\left(V,F\right)$ corresponding to the vertices ($V$) that represent the geometry information and the indexed faces ($F$) of the mesh. Each vertex can be represented as a point $\mathbf{v}_i = (x_i,y_i,z_i) \ \forall \ i = 1,n$ and each centroid of a face as $\mathbf{m}_i = (\mathbf{v}_{i1}+\mathbf{v}_{i2}+\mathbf{v}_{i3})/3 \ \forall \ i = 1,l$. A set of edges ($E$) can be directly derived from $V$ and $F$, which correspond to the connectivity information.

Spectral processing approaches, e.g., \cite{Sorkine2005}, \cite{Karni2000} are based on the fact that smooth geometries should yield spectra, dominated by low frequency components and suggest projecting the Cartesian coordinates $\mathbf{x},\mathbf{y},\mathbf{z}\in \Re^{n\times 1}$ in the basis spanned by the eigenvectors $\mathbf{u}_i$, $i=1,...,c << n$  of the Laplacian operator $\mathbf{L}$ that is calculated as follows:
\begin{eqnarray}
\mathbf{L} &=& \mathbf{D} - \mathbf{C} 
\end{eqnarray} 
where $\mathbf{C} \in  \Re^{n\times n}$ is the weighted connectivity matrix of the mesh with elements:
\begin{equation}
\mathbf{C}_{\left(i,j\right)} = \left\{\begin{array}{ll} \frac{1}{\|\mathbf{v}_i-\mathbf{v}_j\|_2^2} &\left(i,j\right)\in (E)\\
0&otherwise\end{array}\right.
\end{equation}
matrix $\mathbf{D}$ is the diagonal matrix with $\mathbf{D}_{(i,i)} = \left|N(i)\right|$, and $N(i) = \left\{j\left|\right. \left(i,j\right)\in (E)\right\}$ is a set of the immediate neighbors for node $i$. 

%\color{red} I can show experiments with different weighted-adjacency matrices. \color{black}

The weighted adjacency matrix is ideal for emphasizing the coherence between Laplacian matrices of different submeshes by providing geometric information; on the contrary, the binary provides only connectivity information. Eigenvalue decomposition of $\mathbf{L}$ is written as:
\begin{equation}
\mathbf{L} = \mathbf{U}\Lambda\mathbf{U}^T
\end{equation}
where $\Lambda$ is a diagonal matrix consisting of the eigenvalues of $\mathbf{L}$ and $\mathbf{U}=[\mathbf{u}_1,\ldots,\mathbf{u}_n]$ is the matrix with the eigenvectors $\mathbf{u}_i\in \Re^{n\times 1}$ which is needed to generate the spectral coefficients that are essential in providing sparse representations of the raw geometry data \cite{Sorkine2005}.

Similar to classical Fourier transform, the eigenvectors and eigenvalues of the Laplacian matrix $\mathbf{L}$ provide a spectral interpretation of the 3D signal. The eigenvalues $\left\{\lambda_1, \lambda_2, \ldots, \lambda_n\right\}$ can be considered as graph frequencies, and the eigenvectors demonstrate increasing oscillatory behavior as the magnitude of $\lambda_i$ increases \cite{briandavies2001discrete}. The Graph Fourier Transform (GFT) of the vertex coordinates is defined as its projection onto the eigenvectors of the graph, i.e.,  $\bar{\mathbf{v}} = \mathbf{U}^T\mathbf{v}$ and the inverse GFT is given by $\mathbf{v} = \mathbf{U}\bar{\mathbf{v}}$. 

\section{Block-Based Spectral Processing Using OI}
As mentioned earlier, calculating the graph Laplacian eigenvalues of the mesh geometry can become restrictive as the density of the models increases. To overcome this limitation, several approaches suggest processing large meshes into parts \cite{Cayre2003},\cite{lalos2015sparse}. Thus, we assume the original 3D mesh is partitioned into $k$ non-overlapping parts using the MeTiS algorithm described in \cite{Karypis1998}.  To be able to directly apply OI, we require to process sequentially a series of matrices of the same size. To that end, we create overlapped equal-sized submeshes, as described in the paragraphs 6.1 $\&$ 6.3. The evaluation of the eigenvectors of the respective matrix $\mathbf{L}{\left[i\right]} \ \forall i=1,\ldots,k $ requires  $\mathcal{O}(kn^{3}_d)$ floating point operations. To minimize this complexity, we suggest exploiting the coherence between the spectral components of the different submeshes using OI \cite{golub2012matrix}. This assumption is strongly based on the observation that submeshes of the same mesh maintain similar geometric characteristics and connectivity information, which will be further discussed in paragraph 6.4. 

The Orthogonal Iteration is an iterative procedure that can be used to compute the singular vectors corresponding to the dominant singular values of a symmetric, nonnegative definite matrix. Alternatively to the OI, the Lanczos approach could be used. However, the initialization of OI to a starting subspace close to the subspace of interest leads to a very fast solution. This property is efficiently exploited when processing sequential submeshes, leading to a lower total complexity as compared to the complexity of the Lanczos approach. Building on this line of thought we suggest evaluating the $c$ eigenvectors corresponding to the $c$ lowest eigenvalue of $\mathbf{L}{\left[i\right]}$ each submesh $i$, $\mathbf{U}_c \left[i\right]=\left[ \mathbf{u}_1,\ldots, \mathbf{u}_c \right] \in \Re^{n_d\times c}$ according to Algorithm 1,
\begin{algorithm}[h]
	$\mathbf{U}(0)\leftarrow~\mathbf{U}_{c}[i-1];$\\
	\For{$t\leftarrow 1$ \KwTo $t_{max}$}{
		$\mathbf{U}(t)\leftarrow~\textbf{Onorm}(\mathbf{R}^z_i\mathbf{U}(t-1));$\\
	}
	$\mathbf{U}_{c}[i]\leftarrow~\mathbf{U}(t);$
	\caption{Orthogonal Iteration (\textbf{OI}) update process for each submesh $i$   }\label{al:pipeline_AOI}
\end{algorithm}
\noindent where $\mathbf{R}_i=\left(\mathbf{L}{\left[i\right]}+\delta \mathbf{I}\right)^{-1}$ and $\delta$ is a small positive scalar value that ensures positive definiteness of $\mathbf{R}_i$. Matrix $\mathbf{I}$ is the identity matrix of size $n_d\times n_d$. At this point it should be noted that the projected coefficients $\mathbf{R}^z_i\mathbf{U}(t-1)$ are estimated very efficiently using sparse linear system solvers \cite{Sorkine2005}. Depending on the choice of power value $z$, we obtain alternative iterative algorithms with different convergence properties. The convergence rate of OI depends on $| \lambda_{c+1} /\ \lambda_c |^z$ where $\lambda_{c+1}$ is the $(c+1)$-st largest eigenvalue of $\mathbf{R}_i$ \cite{Zhangthesis}.

To preserve orthonormality, it is important that the initial subspace $\mathbf{\widetilde{U}}_c\left[0\right]$ is orthonormal. For that reason, $\mathbf{\widetilde{U}}_c\left[0\right]$ is estimated by applying SVD directly on the first selected submesh\footnote{Please note that the selection of the initial submesh does not affect the transient behavior of the algorithm}, while the following subspaces $\mathbf{\widetilde{U}}_c\left[i\right]$, $i=2,\ldots,k$ are adjusted using Algorithm 1.

The initial submesh is selected in a random order and the subsequent ones are processed in a topologically sorted order.

The orthonormalization of the estimated subspace can be performed using a number of different choices \cite{hua2004asymptotical} that affect both complexity and performance. The most widely adopted are the Householder Reflections (HR), Gram-Schmidt (GS) and Modified GramSchmidt (MGS) methods. Although, the aforementioned variants exhibit different properties related to the numerical stability and computational complexity, the $\textbf{Onorm}(\cdot)$ step is performed as follows: 
\begin{align}
\mathbf{R}^z\left[i\right]\mathbf{\widetilde{U}}_c[i] &\Rightarrow \mathbf{Q}_{qr}\left[i\right]\mathbf{R}_{qr}\left[i\right]\nonumber\\
\mathbf{\widetilde{U}}_c\left[i\right] &= \mathbf{\widetilde{Q}}_{qr}\left[i\right] = \left[\mathbf{Q}_{qr}\left[i\right]_{(:,1)},\ldots, \mathbf{Q}_{qr}\left[i\right]_{(:,c)}\right]
\end{align}
where matrix $\mathbf{\widetilde{Q}}_{qr}\left[i\right]$ is evaluated by applying $c$ sequential HR reflections. Therefore, $\mathbf{\widetilde{Q}}_{qr}\left[ i \right]$ is the submatrix that corresponds to the first $c$ columns of: 
\begin{align}
\mathbf{Q}_{qr} \left[ i \right]=  \mathbf{H_1}^T \cdot \mathbf{H_2}^T \cdot \ldots \cdot \mathbf{H_c}^T 
\label{HS1}
\end{align}

\section{Dynamic OI for Stable Reconstruction}

In application scenarios where the original mesh is known, we propose a flexible solution that automatically identifies the optimal subspace size $c$ that satisfies a specific reconstruction quality criterion. This novel extension can be used for improving the reconstruction quality in special cases where the coherence between submeshes is not strong enough e.g. different density, difference in geometry. The identification is performed sequentially, based on user defined thresholds, that determine the lower and higher ''acceptable'' quality of the reconstructed submeshes at the decoder side. In practical scenarios it is reasonable to assume that the feature vectors $\mathbf{E}[i] = \mathbf{U}^T_c[i]\mathbf{v}[i]$ of each block $\mathbf{v}[i]$ ''live'' in subspaces $\mathbf{U}_c[i]$ of different sizes. The subspace size $c_i$ of the incoming data block $\mathbf{v}\left[i\right]$, should be carefully selected so that the relevant submesh vertices are identified with the minimum loss of information. To quantify this loss at each iteration $t$, we suggest using the $l_2$-norm of the following mean residual vector:
\begin{equation}
\mathbf{e}(t) =  \sum_{j \in \left\{x,y,z\right\}} \left(\mathbf{v}_j\left[i\right]-\mathbf{U}_c\left[i\right]\mathbf{U}_c^T\left[i\right]\mathbf{v}_j\left[i\right]\right)
\label{eq:residual}
\end{equation}
where each $\mathbf{v}_j\left[i\right]$, $\forall \left\{x,y,z\right\}$ correspond to the $n_d\times 1$ vector with the $x$,$y$ and $z$ coordinates of the submesh $i$ vertices. When the $l_2$-norm value of this metric is below a given threshold $\| \mathbf{e}(t) \|_2 < \epsilon_h $ the loss of information during the spectral processing steps is not easily perceived. To reduce the residual error $\mathbf{e}(t)$, we suggest adding one normalized column in the estimated subspace $\mathbf{U}_c(t) = [\mathbf{U}_c(t-1) \ \ \mathbf{e}(t-1)/\| \mathbf{e}(t-1) \|_2]$ and then perform orthonormalization, e.g.,
%\begin{equation}
%e(t-1) =  \sum_{j=1}^{n}(r_{ij}-\mathbf{U}_c(t-1)\mathbf{U}_c^T(t-1)r_{ij} )
%\label{eq:newresidual}
%\end{equation}
\begin{equation}
\mathbf{U}_c(t) =   Onorm\left \{\mathbf{R}^z[i] [\mathbf{U}_c(t-1) \ \ \frac{\mathbf{e}(t-1)}{\| \mathbf{e}(t-1) \|_2}]\right\}
\label{eq:newU}
\end{equation}

Similarly, when the value of the reconstruction quality metric is less than a user determined lower bound  $\epsilon_l$ the subspace size is decreased by 1 by simply selecting the first \(c_i - 1\) columns of $\mathbf{U}_c(t)$. This procedure is repeated until the value of the metric lies within the range  $(\epsilon_l,\epsilon_h)$, allowing the user to easily trade the reconstruction quality with the computational complexity. To summarize, Algorithm~\ref{al:pipeline} presents the steps of the dynamic OI approach.

\begin{algorithm}
	$\mathbf{U}(0)=\mathbf{U}_{i-1}$;  \\
	$c_i \leftarrow (i > 0) \ ? \ c_{i-1} : c$; \\
	\For{$t = 1,2,...$}{
		$\mathbf{U}(t) = [u_1 \dots u_{c_t}] = \mathbf{Onorm}(\mathbf{R}_i^z U(t-1) )$; \\
		$\mathbf{e}(t) =  \sum_{j \in \left\{x,y,z\right\}} \left(\mathbf{v}_{ij}-\mathbf{U}(t)\mathbf{U}(t)^T\mathbf{v}_{ij}\right)$; \\
		\uIf{$\| \mathbf{e}(t) \|_2 < \epsilon_l$}{$\mathbf{U}(t) = [\mathbf{U}(t-1) \ \ \frac{\mathbf{e}(t)}{\| \mathbf{e}(t) \|_2}]; \ c_i \leftarrow c_i + 1$;
		}\uElseIf{$\| \mathbf{e}(t) \|_2 > \epsilon_h$}{$\mathbf{U}(t) = [u_1 \dots u_{c_t}]; \ c_i \leftarrow c_i - 1$;
		}\Else {break;} 
		
	} 
	$\mathbf{U}_i = \mathbf{U}(t)$;
	\caption{Dynamic OI (\textbf{DOI}) applied in any $i$ submesh} \label{al:pipeline}
\end{algorithm}

\section{Ideal Mesh Segmentation and Submeshes Properties}

In this section, we study the impact of the submesh size to the reconstruction quality and the execution time. Additionally, we present the methodology that we follow for the final reconstruction of the mesh meaning that the submeshes are overlapped and some points appear in more that one submesh. The section is concluded with some experimental results confirming the validity of the assumption about the spatial coherence between submeshes of the same meshes.

\subsection{Weighted Average for Mesh Reconstruction and Guarantees of a Smooth Transition}

\begin{figure}
	\centering
	\includegraphics[width=0.85\linewidth]{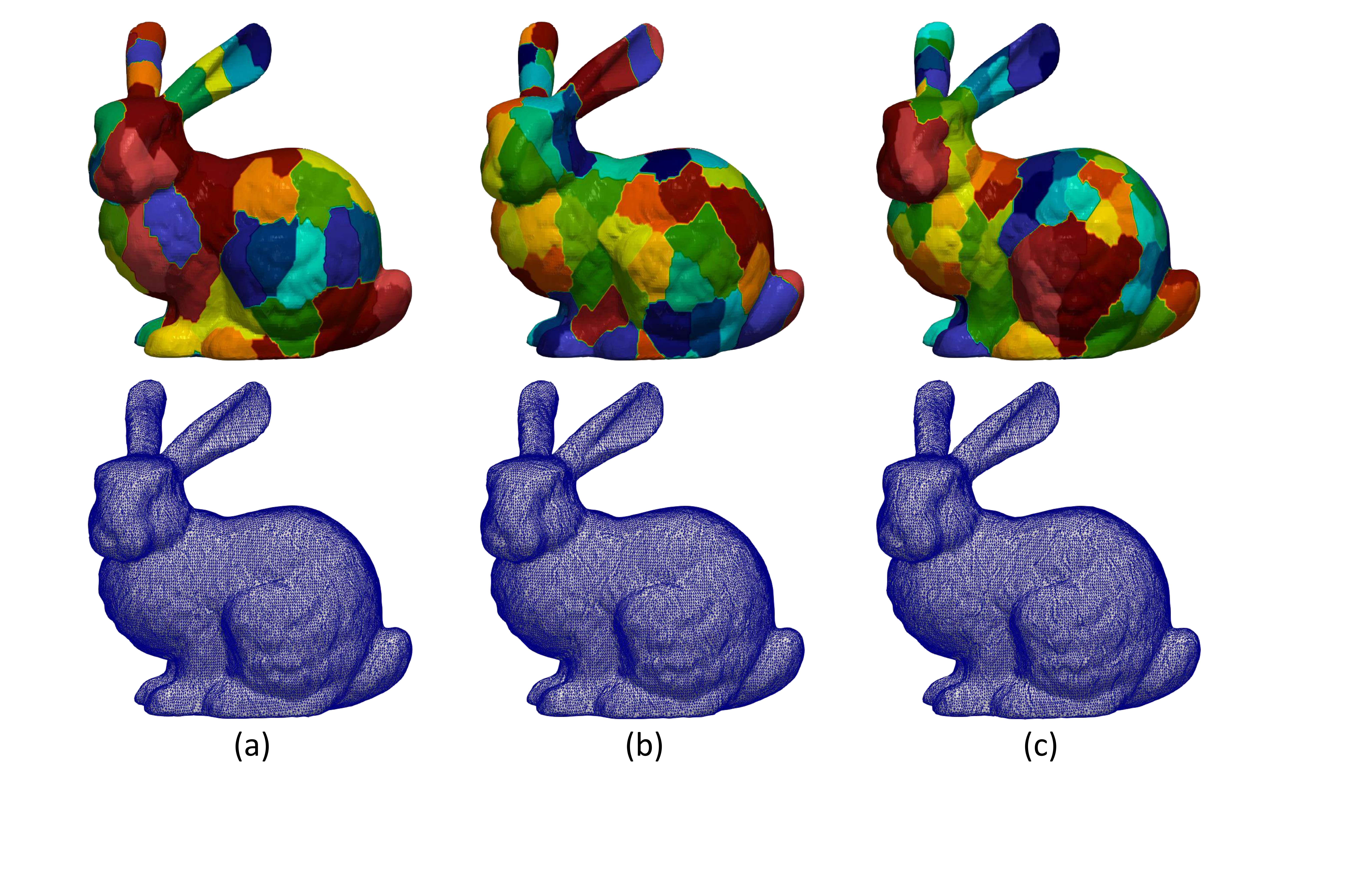}
	\caption{[\textbf{First line}] Segmentation of bunny model using MeTis algorithm in (a) 70, (b) 100 and (c) 200 parts. [\textbf{Second line}] The corresponding reconstructed models without applying overlapping process (edge effect is apparent).}\label{fig:overlapped_bunny}
\end{figure}

Processing of a single mesh in parts usually results in a loss of reconstruction quality that is attributed to the dislocation of the vertices that lie on the edges of each submesh (edge points). These phenomena, also known as edge effects (see Fig. \ref{fig:overlapped_bunny}), can be mitigated by processing overlapped submeshes \cite{Cayre2003},\cite{lalos2015sparse},\cite{8283576}. Therefore each submesh is extended with the neighbors of the boundary nodes of adjacent submeshes consisting in total of $n_d$ nodes. This operation reduces the error introduced when the number of submeshes is increased and additionally creates equal-sized submeshes which are necessary for the proceeding of the OI. In Fig. \ref{fig:overlapped_bunny}, we present different segmentation scenarios using MeTis algorithm. Inspecting the second line of this figure, which presents the reconstructed model highlighting the edges of the triangles, it is apparent that the more the parts of the segmentation are, the more apparent the edge effect is.

\begin{figure}
	\centering
	\mbox{} 
	\includegraphics[width=0.85\linewidth]{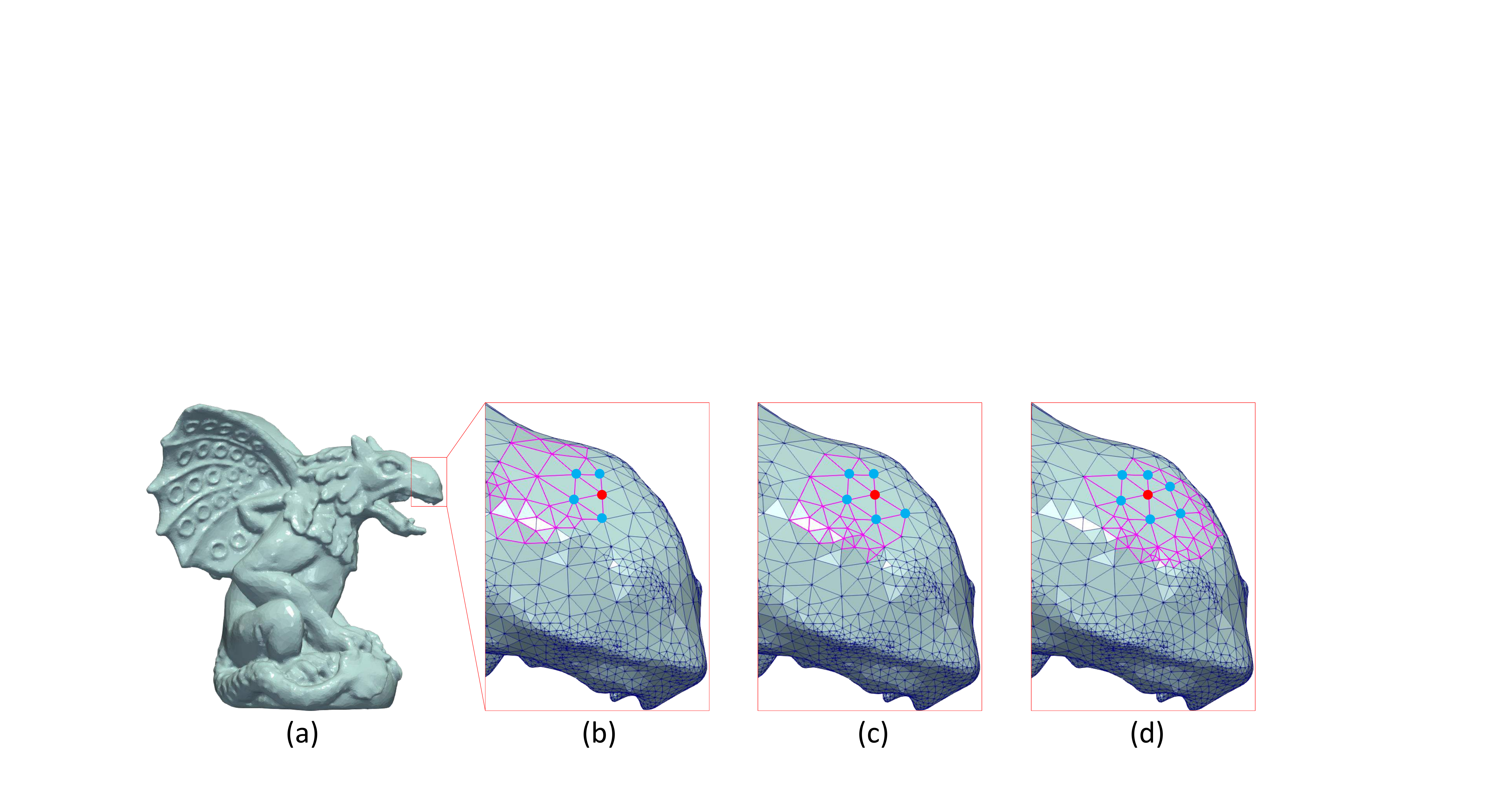}
	\mbox{}
	\caption{The red point has different degree in each submesh, the corresponding weights are: (a) w = 4, (b) w = 5, (c) w = 6 (Gargoyle model). }\label{fig:gargoyle_degree}
\end{figure}

The edge effect is attributed to missing neighbors inevitably caused by the mesh segmentation. Missing neighbors means missing connectivity which resulting in missing entries in the graph Laplacian matrix. However, an efficient way to deal effectively with this limitation is to combine the reconstructed geometry of the overlapped parts. The weights that are assigned to each point are proportional to the degree of the node (e.g., number of neighbors) in the corresponding submesh. Overlapping ensures that each vertex will participate in more than one submesh, and thus the probability of having the same degree (in at least one of them) significantly increases. In Fig. \ref{fig:gargoyle_degree}, we present an example showing the weights assigned to a point (highlighted in red) that participates in three overlapped submeshes. The steps that are followed for the estimation of the weighted average coordinates of the overlapped points, are presented in Algorithm \ref{al:Weighted_average}. 

\begin{algorithm}
	\For{$i = 1,..,n$}{ 
		Find the $p_i \ge 1$ overlapped submeshes in which the $i^{th}$ point appears;\\
		Set the indices of these $p_i$ submeshes in a vector $\mathbf{q}_{i} \in \Re^{p_i\times 1}$;  \\
		$\mathbf{sumv}_{i} = [{sumx}_{i}, \ {sumy}_{i}, \ {sumz}_{i}] = [0, \ 0, \ 0]$; \\
		${sumw}_{i} = 0$; \\
		\For{$j  \in  \mathbf{q}_{i} $}{
			Find the degree $w_{ij}$ of $i^{th}$ point $\mathbf{v}_i = [x_i, \ y_i, \ z_i] $ in the $j^{th}$ submesh; \\
			$\mathbf{sumv}_{i} = \mathbf{sumv}_{i} + w_{ij}\mathbf{v}_{ij}$; \\
			${sumw}_{i} = {sumw}_{i}+w_{ij}$;\\
		}
		$\tilde{\mathbf{v}}_{ij} = \frac{ \mathbf{sumv}_{i} } { {sumw}_{i} }$;
	}
	\caption{Weighted average process for the reconstruction of a mesh} \label{al:Weighted_average}
\end{algorithm}

Additionally, we investigate whether the segmentation and the processing of the overlapped patches guarantee the smooth transition in different cases where edge points belong to flat or sharp areas. At this point it should be mentioned that, the edge points could be part of edges, corners or flat areas. In the following, we present results showing that the way we treat the edge points guarantees, in all the aforementioned cases, a smooth transition successfully mitigating the edge effects. 

\begin{figure}
	\centering
	\mbox{} 
	\includegraphics[width=0.85\linewidth]{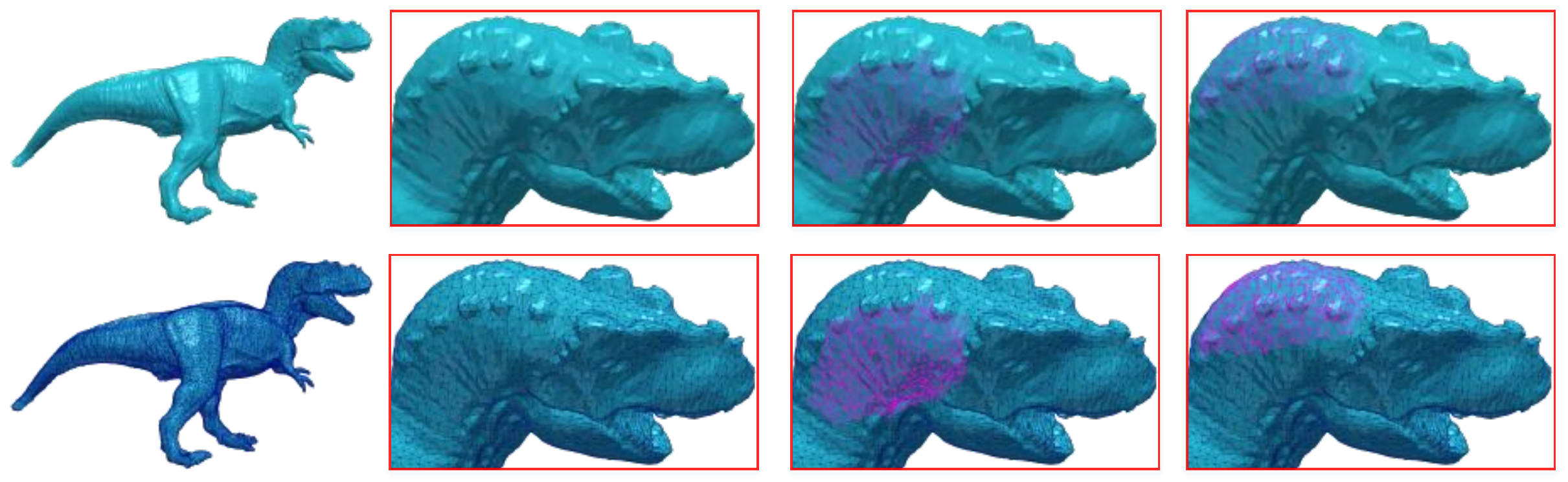}
	\mbox{}
	\caption{Overlapped parts means that each boundary point belongs to more than one part and its degree may vary significantly between different parts. }\label{fig:tyra_overalapped}
\end{figure}

The process starts using the MeTis algorithm for the identification of the initial parts. Then each part is extended, using the neighbors of the boundary nodes that belong to adjacent parts until all of them has the same predefined size. Consequently, each boundary point participates in more than one segments. The weights that are assigned to each point, which participates in more than one parts, represents its degree (i.e, the number of connected neighbors) in the specific part (see Fig. \ref{fig:tyra_overalapped}). The final position of an edge point is evaluated using the weighted average approach as mentioned above.  

We show the distribution of error in the internal and the boundary points of each submesh. For this specific study we consider 3 different cases that are described below:
\begin{itemize}
	\item \textbf{Non Overlapping case}, where each node participates in only one part. 
	\item \textbf{Overlapping case}, where each part is extended using the neighbors of the boundary nodes that belong to adjacent parts. Thus, each boundary point participates in more than one parts, which are reconstructed individually. The final position of a boundary point is evaluated using the simple average of the reconstructed positions. 
	\item \textbf{Weighted Overlapping case}, where each part is extended using the neighbors of the boundary nodes that belong to adjacent parts and the final position of a boundary point is evaluated using a weighted average.  The weights assigned to each point that participates in more than one parts, represent its degree (i.e, the number of its neighbors) in the specific part.
\end{itemize}

\begin{figure}[H]
	\centering
	\mbox{} 
	\includegraphics[width=1\linewidth]{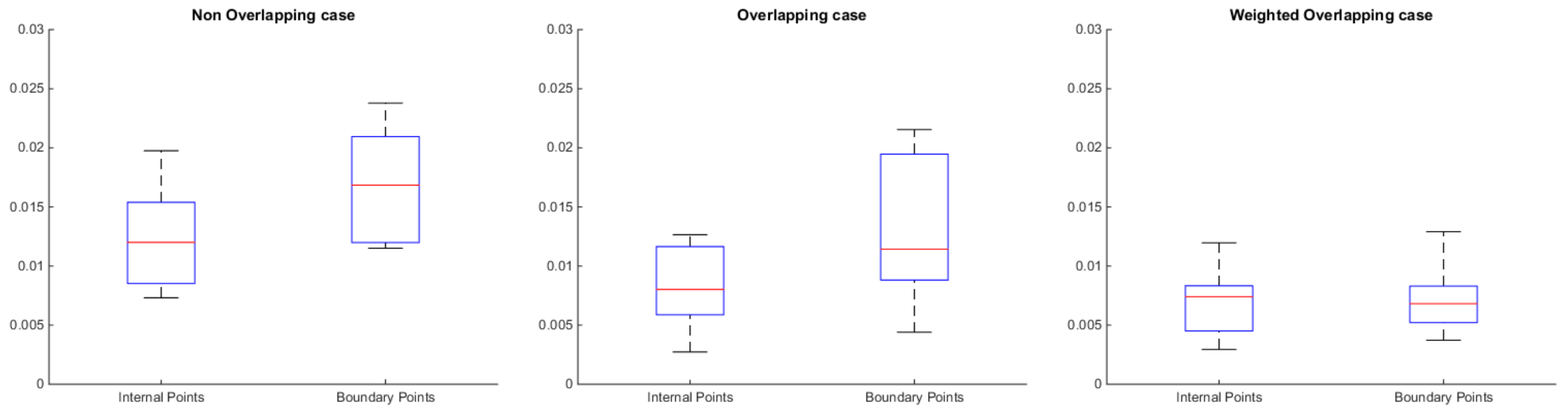}
	\mbox{}
	\caption{Standard deviation of the reconstructed error in the internal and the boundary points of each submesh for each one of the aforementioned cases. }\label{fig:case_boxplot}
\end{figure}

The standard deviation of the reconstructed error in the internal and the boundary points of each submesh for each one of the aforementioned cases is provided in Fig. \ref{fig:case_boxplot}. On each box, the central mark is the median, the edges of the box are the $25^{th}$ and $75^{th}$ percentiles, and the whiskers extend to the most extreme data points that are not considered outliers. By inspecting this figure, it can be clearly shown that the weighting scheme guarantees a smooth transition, since the distribution of error in the internal and boundary points has almost identical characteristics, significantly outperforming the other two cases.

\begin{figure}
	\centering
	\mbox{} 
	\includegraphics[width=0.8\linewidth]{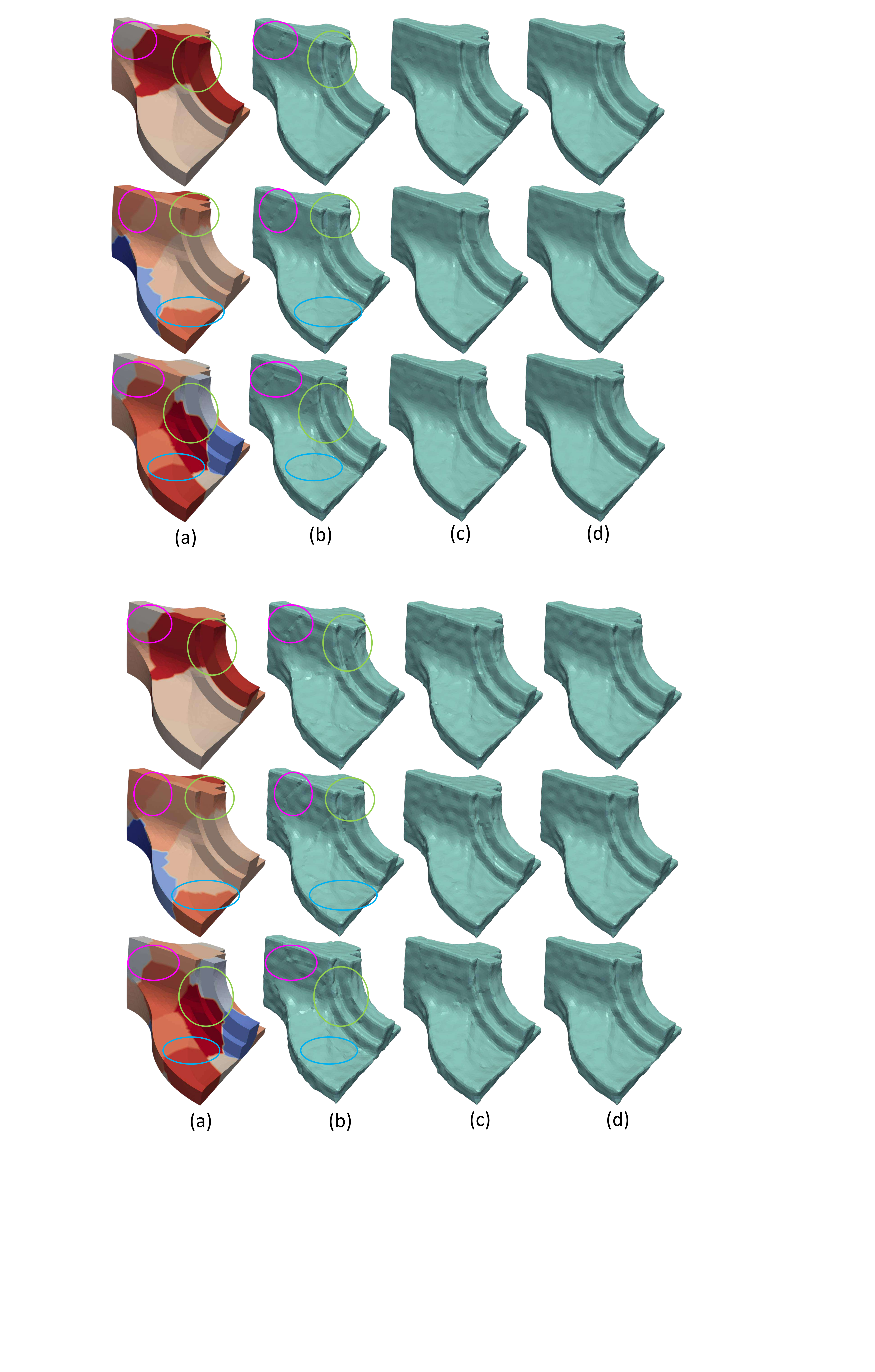}
	\mbox{}
	\caption{(a) The model separated in different number of parts (10 , 15 and 20 respectively). Additionally, indicative areas have been selected where two or more submeshes are connected, (b) \textbf{Non Overlapping case}, the edge effect is apparent in areas where submeshes are connected, (c) \textbf{Overlapping case}, the edge effect have been mitigated but have not been eliminated yet. The bigger the number of the partitioning the more intense the problem of the effect, (d) \textbf{Weighted Overlapping case}, the results seem to be independent and unaffected of the partitioning (Fandisk $\sigma^2 = 0.2$). }\label{fig:cropped_detailed2}
\end{figure}

Similar conclusions can be also perceived by observing the Fig. \ref{fig:cropped_detailed2}. In this figure, the results of a coarse denoising step are presented after partitioning Fandisk model in a different number of submeshes (10 , 15 and 20 respectively). It is obvious that the error on the boundary nodes is minimized in the weighted average case, while the segmentation effects are very noticeable in the other two cases.

\subsection{Number of Submeshes}% Affecting the Denoising Results}
The ideal selected number of submeshes depends on the total number of points of the mesh. Large submeshes create large matrices increasing significantly the processing time since the number of edge points increases. On the other hand, using small submeshes the final results are negatively affected by the edge effects. Table \ref{table:different_part_bunny} shows how the number of segments affects the metric of Mean Normal Difference (MND) for both averaging cases (simple and weighted average), where MND represents the average distance from the resulting mesh normals to the ground truth mesh surface.

\begin{table}
	\centering
	\begin{tabular}{|c|c|c|c|}
		\hline
		\textbf{ \thead{Number of \\ submeshes} } & \textbf{ \thead{Number of \\ Vertices per \\ Segment}} & \textbf{ \thead{MND \\ using \\ simple \\ average}} & \textbf{ \thead{MND \\ using \\ weighted \\ average}}\\
		\hline
		25 & 1392 & 0.0921 & 0.0915 \\
		\hline
		\textbf{40} & $\sim$ \textbf{870} & \textbf{0.0931} & \textbf{0.0925} \\
		\hline
		\textbf{50} & $\sim$ \textbf{696} & \textbf{0.0941} & \textbf{0.0934} \\
		\hline
		\textbf{70} & $\sim$ \textbf{497} & \textbf{0.0960} & \textbf{0.0952} \\
		\hline
		100 & $\sim$ 348 & 0.0988 & 0.0980 \\
		\hline
		200 & $\sim$ 174 & 0.1039 &  0.1028 \\
		\hline
		500 & $\sim$ 69 & 0.1163 & 0.1150 \\
		\hline
	\end{tabular}
    \vspace{0.3em}
	\caption{Mean Normal Difference using different number of segments (Bunny Model with 34817 vertices). We also compare the mean normal difference by using normal average and weighted average based on the number of the connected vertices.} \label{table:different_part_bunny}
\end{table}

\begin{figure}
	\centering
	\mbox{} 
	\includegraphics[width=0.9\linewidth]{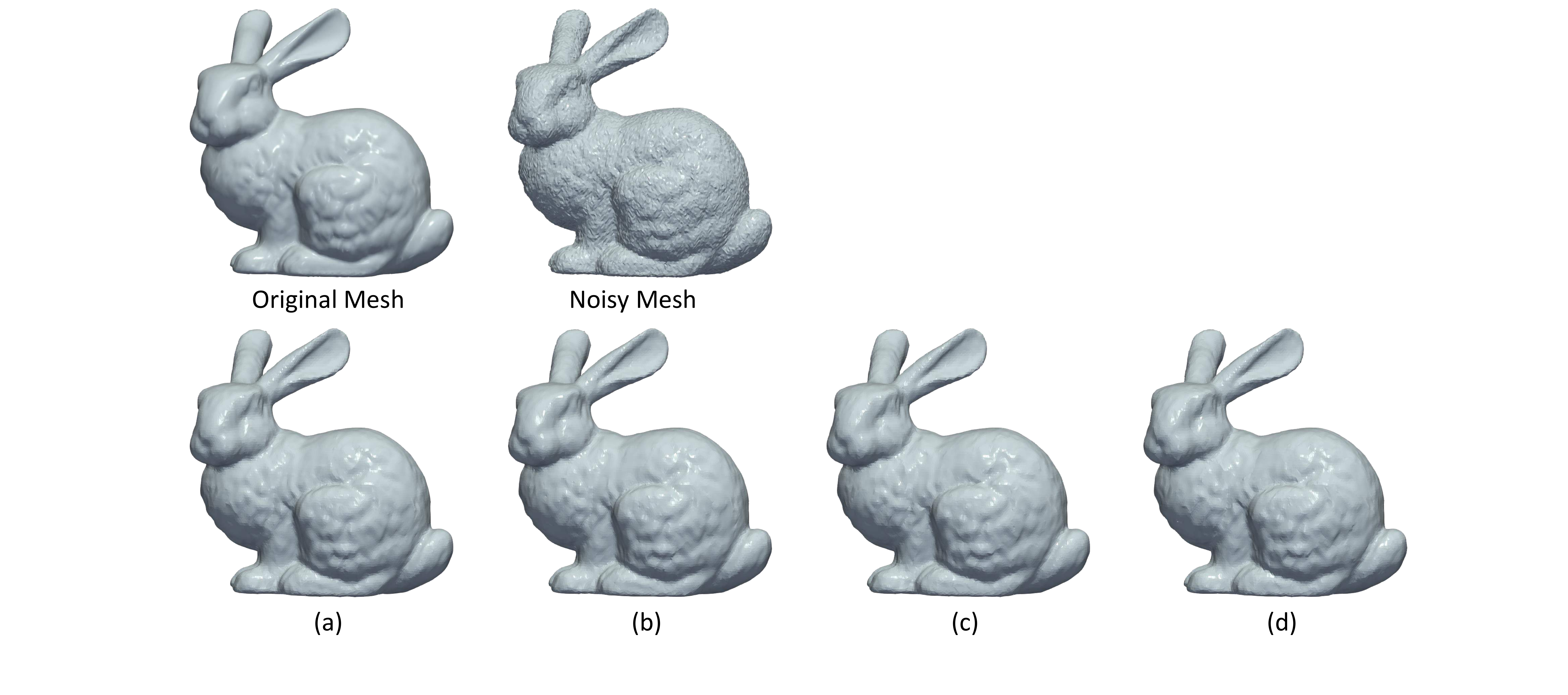}
	\mbox{}
	\caption{\textbf{[First line]} Original and Noisy mesh. \textbf{[Second line]} Coarse denoising meshes separated by Metis in (a) 25 submeshes (b) 50 submeshes (c) 70 submeshes (d) 100 submeshes.}\label{fig:different_parts_bunny}
\end{figure}

In Fig. \ref{fig:different_parts_bunny}, the results of coarse smoothing, using a different number of segments, are also presented. As we can observe, there is no remarkable visual difference between the reconstructed models. Additionally, if we consider the fact that these results could be further improved by the use of a fine denoising step then the number of segments is not a critical factor.

\subsection{Size of Overlapped Submeshes}% Affecting the Denoising Results}

The real motivation behind the processing in parts, is strongly supported by the existence of a great amount of state-of-the-art applications in which large 3D models cannot be scanned at once using portable 3D scanners. As a result, the output of the sequential scanning would be a sequence of submeshes that arrive sequentially in time. An extensive evaluation study carried out using different overlapped sizes (Tables \ref{table:julio_overlapped50}, \ref{table:julio_overlapped70}, \ref{table:julio_overlapped100}) showed that the reconstruction quality is strongly affected by the size of the submeshes themselves rather than the number of overlapped vertices.

\begin{figure}
	\centering
	\mbox{} 
	\includegraphics[width=0.75\linewidth]{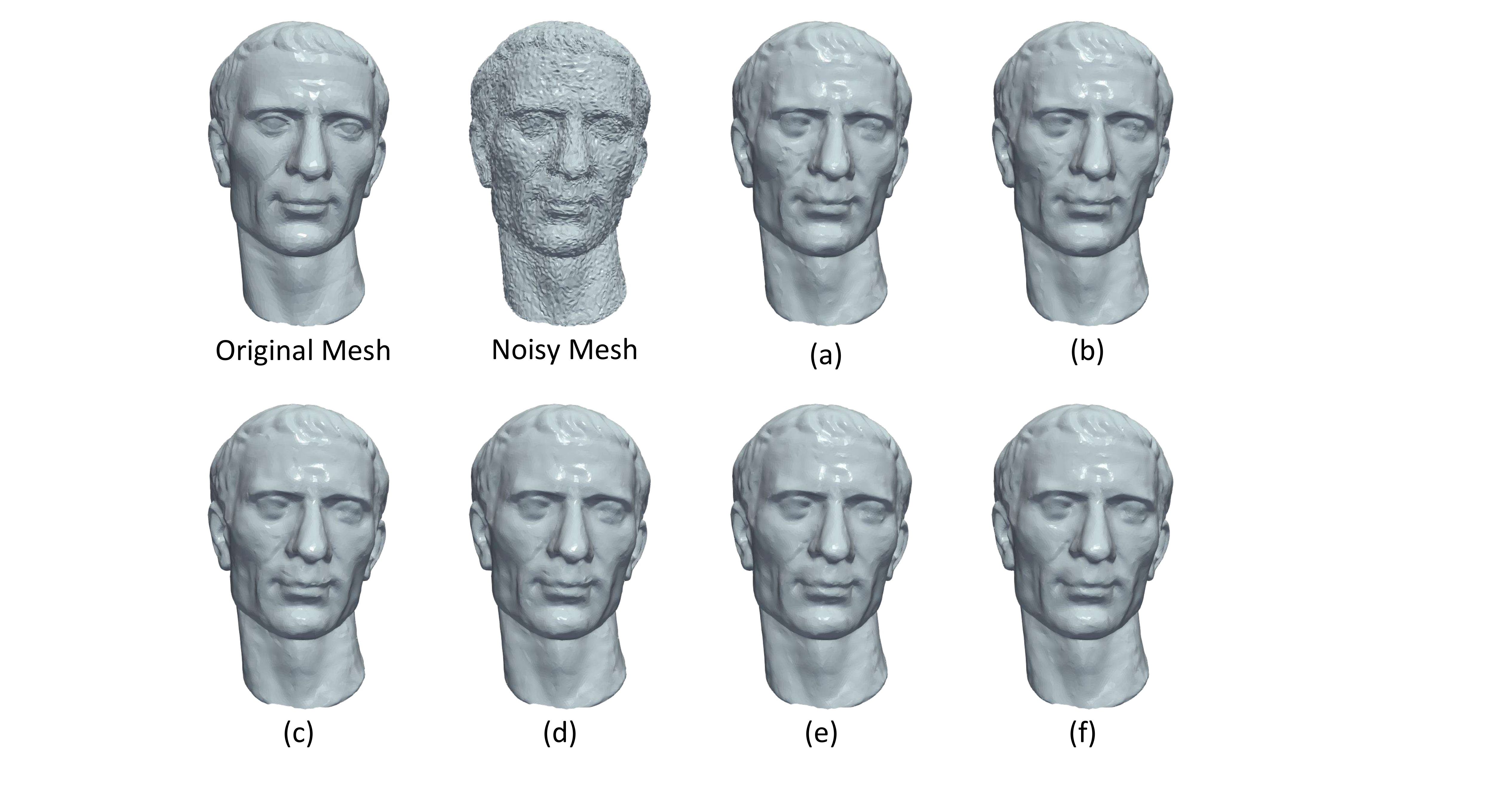}
	\mbox{}
	\caption{Coarse denoising meshes with 70 equal-sized overlapped submeshes consisting of (a) 532 vertices (max) (b) 558 vertices (1.05 $\cdot$ max) (c) 585 vertices (1.10 $\cdot$ max) (d) 611 vertices (1.15 $\cdot$ max) (e) 638 vertices (1.20 $\cdot$ max) (f) 665 vertices (1.25 $\cdot$ max).}\label{fig:julio_overlapped}
\end{figure}

Regarding the ideal size of the overlapped patches, we investigated the effect of using different sizes of overlapped submeshes in a range from 5\% to 25\% of the maximum submeshes length, in the quality of the reconstructed model. More specifically, as shown in Tables \ref{table:julio_overlapped50} - \ref{table:julio_overlapped100} and in Fig. \ref{fig:julio_overlapped}, the mean normal difference and the visual smoothed results have not significant differences between the different case studies, especially for percentages up to 10\% of the max segment. Additionally, if we consider the fact that this process takes place in the coarse denoising step we can conceive the negligible contribution of the overlapped submeshes size to the final denoising results. 

By inspecting the results, we can definitely state that the number and size of segments are much more important than the size of the overlapped patches. The overlapping process mainly contributes in the case of on-the-edge points helping for a more accurate estimation of their position by creating full-connected points. A sufficient overlapping size corresponds to the 15\% of the total points in the submesh. 

Fig. \ref{fig:julio_overlapped} illustrates the reconstruction results of the coarse denoising step using 70 overlapped submeshes consisting of a different number of vertices in each case. As we can observe, in cases where the number of overlapping vertices is higher than $15\%$ of the total number of submesh points then the reconstructed results are almost identical with the $15\%$ case.

\begin{table}
	\centering
	\begin{tabular}{|c|c|c|c|}
		\hline
		\textbf{ \thead{Type of \\ overlapping} }& \textbf{ \thead{Number of \\ Vertices \\ per Segment} } & \textbf{ \thead{Coarse \\ Denoising \\ MND} } & 
		\textbf{\thead{Fine \\ Denoising \\ MND} }\\
		\hline
		max & 741 &  0.1229 & 0.1167 \\
		\hline
		1.05 $\cdot$ max & 778 & 0.1207 & 0.1166 \\
		\hline
		1.10 $\cdot $ max & 815 & 0.1188 & 0.1163 \\
		\hline
		1.15 $\cdot$ max & 852 & 0.1172 & 0.1160 \\
		\hline
		1.20 $\cdot$ max & 889 & 0.1160 & 0.1159 \\
		\hline
		1.25 $\cdot$ max & 926 & 0.1159 & 0.1158 \\
		\hline
	\end{tabular}
    \vspace{0.3em}
	\caption{Mean normal difference using different size of equal-sized overlapped submeshes (Julio Model with 36201 vertices 50 segments)}
	\label{table:julio_overlapped50}
\end{table}

\begin{table}
	\centering
	\begin{tabular}{|c|c|c|c|}
		\hline
		\textbf{ \thead{Type of \\ overlapping} }& \textbf{ \thead{Number of \\ Vertices \\ per Segment} } & \textbf{ \thead{Coarse \\ Denoising \\ MND} } & 
		\textbf{\thead{Fine \\ Denoising \\ MND} }\\
		\hline
		max & 532 & 0.1248 & 0.1176 \\
		\hline
		1.05 $\cdot$ max & 558 & 0.1228 & 0.1173 \\
		\hline
		1.10 $\cdot $ max & 585 & 0.1203 &  0.1172 \\
		\hline
		1.15 $\cdot$ max & 611 & 0.1188 &  0.1169 \\
		\hline
		1.20 $\cdot$ max & 638 & 0.1174 & 0.1169 \\
		\hline
		1.25 $\cdot$ max & 665 & 0.1164 &  0.1164 \\
		\hline
	\end{tabular}
    \vspace{0.3em}
	\caption{Mean normal difference using different size of equal-sized overlapped submeshes (Julio Model with 36201 vertices 70 segments)}
	\label{table:julio_overlapped70}
\end{table}

\begin{table}
	\centering
	\begin{tabular}{|c|c|c|c|}
		\hline
		\textbf{ \thead{Type of \\ overlapping} }& \textbf{ \thead{Number of \\ Vertices \\ per Segment} } & \textbf{ \thead{Coarse \\ Denoising \\ MND} } & 
		\textbf{\thead{Fine \\ Denoising \\ MND} }\\
		\hline
		max & 372 & 0.1276 & 0.1189\\ 
		\hline
		1.05 $\cdot$ max & 390 & 0.1248 & 0.1187\\
		\hline
		1.10 $\cdot $ max & 409 & 0.1228 & 0.1185 \\
		\hline
		1.15 $\cdot$ max & 427 & 0.1208 & 0.1183 \\
		\hline
		1.20 $\cdot$ max & 446 & 0.1184  & 0.1174 \\
		\hline
		1.25 $\cdot$ max & 465 & 0.1175 & 0.1168 \\ 
		\hline
	\end{tabular}
    \vspace{0.3em}
	\caption{Mean normal difference using different size of equal-sized overlapped submeshes (Julio Model with 36201 vertices 100 segments)}
	\label{table:julio_overlapped100}
\end{table}

\subsection{Spatial Coherence Between Submeshes of the Same Mesh}

The previously presented approach, using OI for the estimation of matrices $\mathbf{\widetilde{U}}_c[i] \ \forall \ i = 1,\dots,k$, strongly depends on the assumption that there is a spatial coherence between submeshes of the same mesh. Supposing the correctness of this assumption, the matrix $\mathbf{U}_{c}[i-1]$, which is used for initializing Algorithm \ref{al:pipeline_AOI}, is the best-related approximation meaning that its form is very close to the real solution. The best-provided initialization matrix has as a result a faster convergence, providing at the same time the most reliable results. In this approach, the proposed initialization strategy suggests using as initial estimation the solution of the previous adjacent submesh.
\\\\
At the following, we will study the validity of this assumption via extensive simulations using different models. Our study is based on the observation that the surface's form of a mesh follows the same pattern, which means that neighboring parts of the same mesh have: 

\begin{enumerate}[label=(\roman*)]
	\item Similar connectivity properties (degree and distance). 
	\item Same geometrical characteristics which are shared between connected points (curvature, small-scale features, texture pattern etc.).
\end{enumerate}

\begin{figure}
	\begin{center}
		\includegraphics[width =1\linewidth]{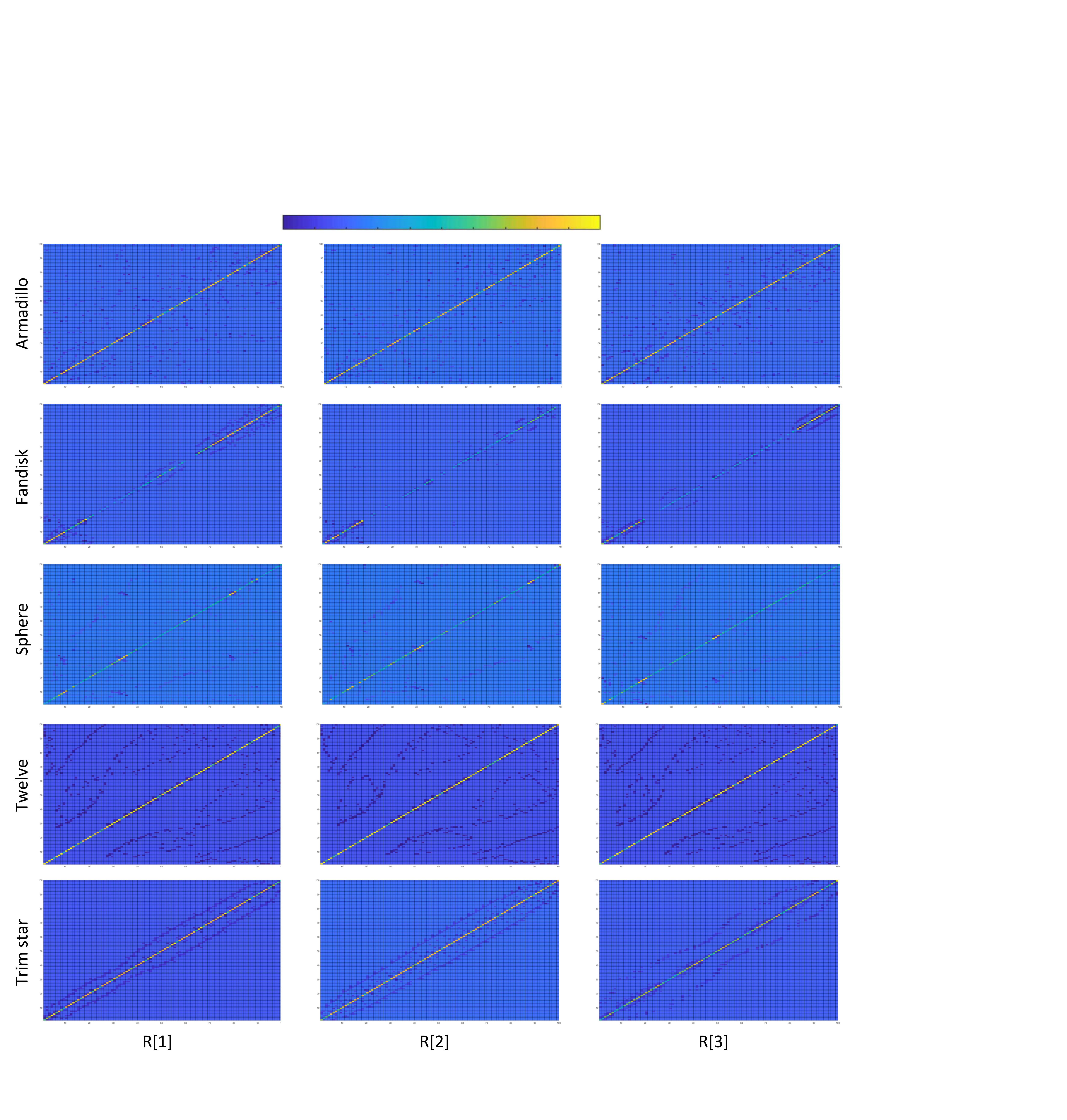}
	\end{center}
	%\vspace{-1em}
	\caption{Laplacian matrices of different submeshes for different models in color based on the values of their cells. It can be easily observed that different submeshes of the same model follow a similar form while they are totally different in comparison with submeshes of different meshes.}
	\label{coherence}
\end{figure}

Fig. \ref{coherence} presents colored images representing the Laplacian matrices $\mathbf{R}[i] \ \forall \ i = 1,2,3$ of different submeshes for several 3D models. Providing an easier comparison between the images, we have created matrices of submeshes with the same size $100 \times 100$ so that $\mathbf{R} \in \Re^{100 \times 100}$. Each pixel $(x,y)$ of an image represents the corresponding color coded value of $\mathbf{R}(x,y)$. Additionally, a color bar is also provided showing the range of colors between the lowest and the highest value of each matrix $\mathbf{R}$, where, the deep blue represents the lowest value of each matrix while the bright yellow represents the highest value. We can observe that different submeshes of the same model follow a similar form while they are totally different in comparison with submeshes of different meshes.

\begin{table}[H]
	\begin{center}
		\begin{tabular}{ | c | c | c | c | c | c|}
			%\hline
			\cline{2-6}
			\multicolumn{1}{c|}{}  & \rule{0pt}{12pt} \textbf{Armadillo}  $\mathbf{\tilde{R}}$ & \textbf{Fandisk} $\mathbf{\tilde{R}}$ & \textbf{Sphere}  $\mathbf{\tilde{R}}$ & \textbf{Trim star}  $\mathbf{\tilde{R}}$ & \textbf{Twelve}  $\mathbf{\tilde{R}}$\\ \hline
			\textbf{Armadillo} $\mathbf{R}[1]$  & \textbf{0.0606} & 13.9720 & 10.0905 &	1.2347& 37.4199 \\ \hline
			\textbf{Fandisk} $\mathbf{R}[1]$ & 15.6144 & \textbf{1.4120} &	11.1582 & 8.4506 & 29.8815 \\ \hline
			\textbf{Sphere} $\mathbf{R}[1]$ & 10.4615 & 11.4700 & \textbf{0.8857} &	3.9065 & 26.4125 \\ \hline
			\textbf{Trim star} $\mathbf{R}[1]$ & 1.3122 & 8.5019 & 3.919 &	\textbf{0.6095} & 29.6996 \\ \hline
			\textbf{Twelve} $\mathbf{R}[1]$ & 37.8481 & 30.2103 & 26.6648 & 30.0618 & \textbf{4.5641} \\ \hline
		\end{tabular}
		\caption { Mean squared error between the $\mathbf{R}[1]$ of different models and the mean $\mathbf{\tilde{R}}$ of each model.}
		\label{table:correl}
	\end{center}
\end{table}

Similar conclusions could be perceived by observing the Table \ref{table:correl}. Each row of this table presents the Mean Squared Error (MSE) estimated by the comparison between the random matrix $\mathbf{R}$ of a model, represented as $\mathbf{R}[1]$, and the mean matrix $\mathbf{\tilde{R}}$ of any other model which appear in Fig. 	\ref{coherence}, including the mean matrix of the same model. This comparison is repeated using different models (other rows of this table). For the shake of simplicity, we used only one random matrix $\mathbf{R}[1]$. However, similar results are extracted using any other random matrix of a model.

\section{Applications}

The primary purpose of this work is the creation of a framework for fast and effective spectral processing of large 3D meshes. In this section we present two case studies: a) compression, b) denoising (both for static and dynamic meshes) where the proposed schema can be applied. 

\subsection{Block-Based Spectral Compression of 3D Meshes}

The spectral compression and reconstruction of static meshes utilize the subspace $\mathbf{\widetilde{U}}_c[i]$ for encoding and decoding the raw geometry data. During the encoding step, the dictionary $\mathbf{\widetilde{U}}_c[i]$ is evaluated, either by direct SVD or by executing a number of OI on $\mathbf{R}^z\left[i\right]$, and is used for providing a compact representation of the Euclidean coordinates of each submesh, e.g. for coordinates $\mathbf{v}_x\left[i\right] \in \Re^{n_d\times 1}$, $\mathbf{E}\left[i\right]=\mathbf{U^T_c}[i] \mathbf{v}\left[i\right]$, where $\mathbf{E}\left[i\right] \in  \Re^{c\times 1}$ and $c<<n_{d_i}$. At the decoder side the original 3D vertices of each submesh are reconstructed from the feature vector $\mathbf{E}\left[i\right]$ and the dictionary $\mathbf{U_c}[i]$ according to : $\tilde{\mathbf{v}}\left[i\right] = \mathbf{U}_c[i]\mathbf{E}\left[i\right]$. Note that the subspace size $c$ remains fixed in the OI case, satisfying fast streaming scenarios, while DOI approach aims at providing high and stable reconstruction accuracy. It is important to mention that the only information transmitted from the sender is the connectivity of the mesh and the $c$ respective spectral coefficients of each block. At the receiver's side, the dictionary $\mathbf{\widetilde{U}}_c[i]$ is evaluated utilizing the connectivity information. For the decoding process, the received spectral coefficients and the dictionary are used to retrieve the original Euclidean coordinates $\mathbf{\hat{x}_i},\mathbf{\hat{y}_i},\mathbf{\hat{z}_i}$, e.g. $\mathbf{\hat{x}_i}= \mathbf{\widetilde{U}}_c[i]\mathbf{s}_{x_i}$. Spectral compression enables aggressive compression ratios, without introducing a significant loss on the visual quality \cite{Karni2000}. 

\subsection{Block-Based Spectral Denoising of 3D Meshes}

\begin{figure}[H]
	\centering
	\includegraphics[width=0.7\linewidth]{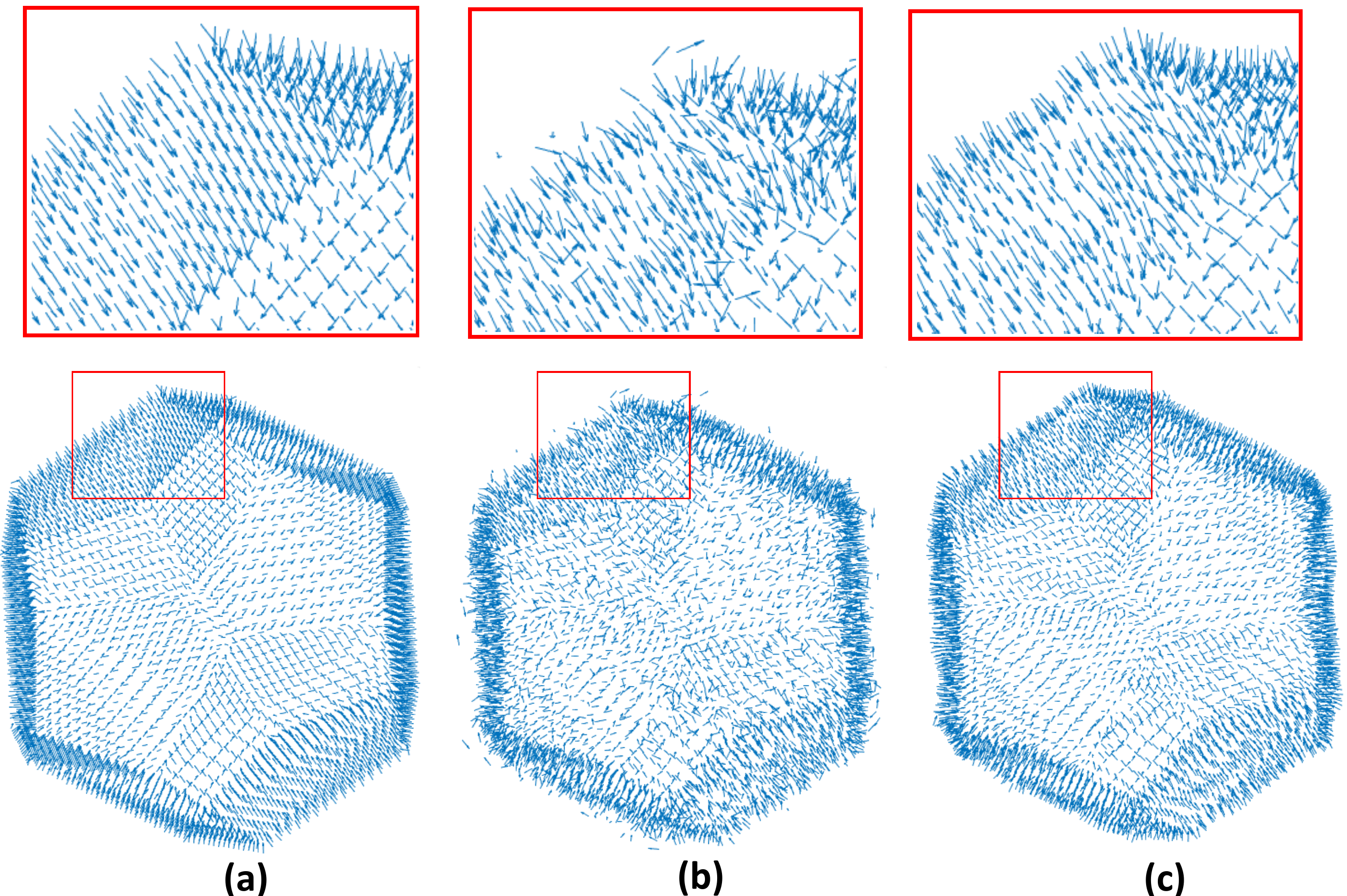}
	\caption{\label{fig:normals_vector}%
		(a) Normals vector of original mesh, (b) Normals vector of noisy mesh, (c) Smoothed normals vector.}
\end{figure}

Bilateral techniques have been used as mesh denoising method in many studies \cite{bilateral}, \cite{bilateralnormal}, \cite{guidedfilter} by iteratively adjusting the face normals and vertices. In this section, we suggest executing a coarse-to-fine spectral denoising method that initially filters out the high spectral frequencies using the aforementioned approach and then performs a fine denoising step using a two stage bilateral technique. The use of the coarse step significantly accelerates the convergence of the fine since it filters out the noise that appears in the higher frequencies, providing a set of normal vectors that are closer to the normal vectors of the original model, as it is clearly shown in the dodecahedron model Fig. \ref{fig:normals_vector}.

We finally show that the fine technique can be also considered as Graph Spectral Processing approach. If we denote with $\hat{\mathbf{v}}[i]=\mathbf{U}_c\mathbf{U}^T_c\mathbf{v}[i]$ the vertices of the coarse denoised $i$ submesh, then each face can be represented by its centroid point $\mathbf{m}_i$, and its outward unit normal:
\begin{align}
\mathbf{\hat{n}}_{m_i} = \frac{(\hat{\mathbf{v}}_{i_{2}}-\hat{\mathbf{v}}_{i_{1}}) \times (\hat{\mathbf{v}}_{i_{3}}-\hat{\mathbf{v}}_{i_{1}}) } {\begin{Vmatrix}(\hat{\mathbf{v}}_{i_{2}}-\hat{\mathbf{v}}_{i_{1}}) \times (\hat{\mathbf{v}}_{i_{3}}-\hat{\mathbf{v}}_{i_{1}})\end{Vmatrix}} \ \forall \ i= 1,n_f \label{eq:ncestimate}
\end{align}
where \(\hat{\mathbf{v}}_{i_{1}}\), \(\hat{\mathbf{v}}_{i_{2}}\), \(\hat{\mathbf{v}}_{i_{3}}\) are the vertices that are related with face \(f_{i}\) and $\mathbf{{\hat{n}}}_{m} = [\mathbf{{\hat{n}}}_{m_1}^T \ \mathbf{{\hat{n}}}_{m_2}^T \cdots \ \mathbf{{\hat{n}}}_{nf}^T] \in \Re^{3n_f \times 1}$.

The bilateral technique estimates the new face normals $\mathbf{n}_i$ using a normal guidance unit vector $\mathbf{g}_i$, which it is calculated as a weighted average of normals in a neighborhood of $i$ is computed by:

\begin{equation}
\hat{\mathbf{n}}_{m_i} = \frac{1}{W_i}\sum_{f_j\in \mathcal{N}_{f_i}} A_j K_s\left(\mathbf{m}_i,\mathbf{m}_j\right) K_r\left(\mathbf{n}_{m_i},\mathbf{n}_{m_j}\right) \mathbf{n}_{m_j}\label{BF}
\end{equation}

\noindent where $ \mathcal{N}_{f_i}$ is the set of faces in a neighborhood of $f_i$, $A_j$ is the area of face $f_j$, $W_i$ is a weight that ensures that $\hat{\mathbf{n}}_{m_i}$ is a unit vector and $K_s$, $K_r$ are the spatial and range Gaussian kernels. More specifically, $K_s$ is monotonically decreasing with respect to the distance of the centroids $\mathbf{m}_i$ and $\mathbf{m}_j$ which lie on the mesh surface, while $K_r$ is monotonically decreasing with the proximity of the guidance normals that lie on the unit sphere:

\begin{eqnarray}
K_s\left(\mathbf{m}_i, \mathbf{m}_j\right) &=& exp\left(-\frac{\left\|\mathbf{m}_i-\mathbf{m}_j\right\|^2}{2\sigma^2_s}\right) \label{Wm}\\
K_r\left(\mathbf{n}_{m_i}, \mathbf{n}_{m_j}\right) &=& exp\left(-\frac{\left\|\mathbf{n}_{m_i}-\mathbf{n}_{m_j}\right\|^2}{2\sigma^2_r}\right)\label{Wn}
\end{eqnarray}
%\begin{align}
%\mathbf{\bar{n}}_{m} =  (\mathbf{D}^{-1} \mathbf{C}_w)^\zeta \mathbf{\hat{n}}_{m}. \label{ni}
%\end{align}

The Bilateral filter output is then used to update the vertex positions in order to match the new normal directions $\mathbf{n}_{m_i}$, according to the iterative scheme proposed in \cite{guidedfilter}. More specifically, the vertex positions $\hat{\mathbf{v}}_{i_1}$, $\hat{\mathbf{v}}_{i_2}$, $\hat{\mathbf{v}}_{i_3}$ of a face $f_i$ are updated in an iterative manner, according to:

\begingroup
\begin{eqnarray}
\hat{\mathbf{v}}^{\left(t+1\right)}_{i_j} &=& \hat{\mathbf{v}}^{\left(t\right)}_{i_j}+\frac{1}{\left|\mathcal{F}_{i_j}\right|}\sum_{z\in \mathcal{F}_{i_j}} \hat{\mathbf{n}}_{m_z} \left[\hat{\mathbf{n}}^T_{m_z}\left(\mathbf{m}^{\left(t\right)}_{i} - \hat{\mathbf{v}}^{\left(t\right)}_{i_j}\right)\right]\label{iter1}\\
\mathbf{m}^{\left(t\right)}_{i}  &=& \left(\hat{\mathbf{v}}^{\left(t\right)}_{i_1}+\hat{\mathbf{v}}^{\left(t\right)}_{i_2}+\hat{\mathbf{v}}^{\left(t\right)}_{i_3}\right)/3\label{iter2}
\end{eqnarray}
\endgroup

where $(t)$ denotes the iteration number, $\mathcal{F}_{i_j}$ is the index set of incident faces for $\hat{\mathbf{v}}_{i_j}$. This iterative process can be considered as a gradient descent process that is executed for minimizing the following energy term across all faces
\begin{equation}
\sum_{z\in \mathcal{F}_{i_j}}\left|\hat{\mathbf{n}}^T_{m_z}\left(\mathbf{m}^{\left(t\right)}_{i} - \hat{\mathbf{v}}^{\left(t\right)}_{i_j}\right)\right|^2,\ j=1,2,3
\end{equation}

This term penalizes displacement perpendicular to the tangent plane defined by the vertex position $\hat{\mathbf{v}}^{\left(t\right)}_{i_j}$ and the local surface normal $\hat{\mathbf{n}}_{m_z}$

\subsubsection*{Bilateral filter as a graph based transform:} Consider an undirected graph $\mathcal{G} = \left(\mathcal{V},\mathcal{E}\right)$ where the nodes $\mathcal{V} = \left\{1,2,\ldots,n\right\}$ are the normals $\mathbf{n}_{m_i}$, associated with the centroids $\mathbf{m}_i$ and the edges $\mathcal{E} = \left\{\left(i,j,c_{ij}\right)\right\}$ capture the similarity between two normals as given by the bilateral weights in Eq. \eqref{Wm}, \eqref{Wn}. The input normals can be considered as a signal defined on this graph $\mathbf{n}_i: \mathcal{V} \rightarrow \Re^{3\times 1}$ where the signal value at each node correspond to the normal vector. Let $\mathbf{C}$ be the adjacency matrix with the bilateral weights and $\mathbf{D} = diag\left\{W_1,\ldots, W_{n_f}\right\}$ the diagonal degree matrix, then Eq. \eqref{BF} can be written as:
\begin{eqnarray}
\hat{\mathbf{n}} &=& \mathbf{D}^{-1}\mathbf{C}\mathbf{n}\nonumber\\
&=& \mathbf{D}^{-1/2}\mathbf{D}^{-1/2}\mathbf{C}\mathbf{D}^{-1/2}\mathbf{D}^{1/2}\mathbf{n}\nonumber\\
\mathbf{D}^{1/2}\hat{\mathbf{n}}&=& \left(\mathbf{I}-\mathbf{L}\right)\mathbf{D}^{1/2}\mathbf{n}\nonumber\\
\mathbf{D}^{1/2}\hat{\mathbf{n}}&=& \underbrace{\mathbf{U}}_{IGFT}\underbrace{\left(\mathbf{I}-\boldsymbol{\Lambda}\right)}_{\begin{array}{c}Spectral\\response\end{array}} \underbrace{\mathbf{U}^T}_{GFT}\mathbf{D}^{1/2}\mathbf{n}
\end{eqnarray}
Thus, it is clearly shown that the Bilateral filter can be considered as a frequency selective graph transform with a spectral response that corresponds to a linear decaying function, meaning that it tries to preserve the low frequency components and attenuate the high frequency ones.

\subsection{Block-Based Spectral Denoising of 3D Dynamic Mesh}

In previous sections, we mentioned that the Laplacian matrices of submeshes, representing parts of the same 3D model, have similar form confirming the existence of spatial coherence. As we presented, we can take advantage of this property implementing a more efficient OI process providing both faster convergence and more accurate results.

However, the advantages of this approach could be better highlighted in the dynamic case. A dynamic mesh consists of s frames/meshes which are shared the same connectivity with each other. Apparently, the Laplacian matrices of corresponding submeshes $\mathbf{R}[i] \ \forall \ i = 1, ... , k$ are preserved the same, without changing, by frame to frame (e.g. $\mathbf{R}_1[1] = \mathbf{R}_2[1] = \cdots = \mathbf{R}_s[1])$, where  $\mathbf{R}_j[i]$ represents the Laplacian matrix of the $i^{th}$ submesh of the $j^{th}$ frame. 

\begin{figure}[H]
	\centering
	\mbox{} 
	\includegraphics[width=0.95\linewidth]{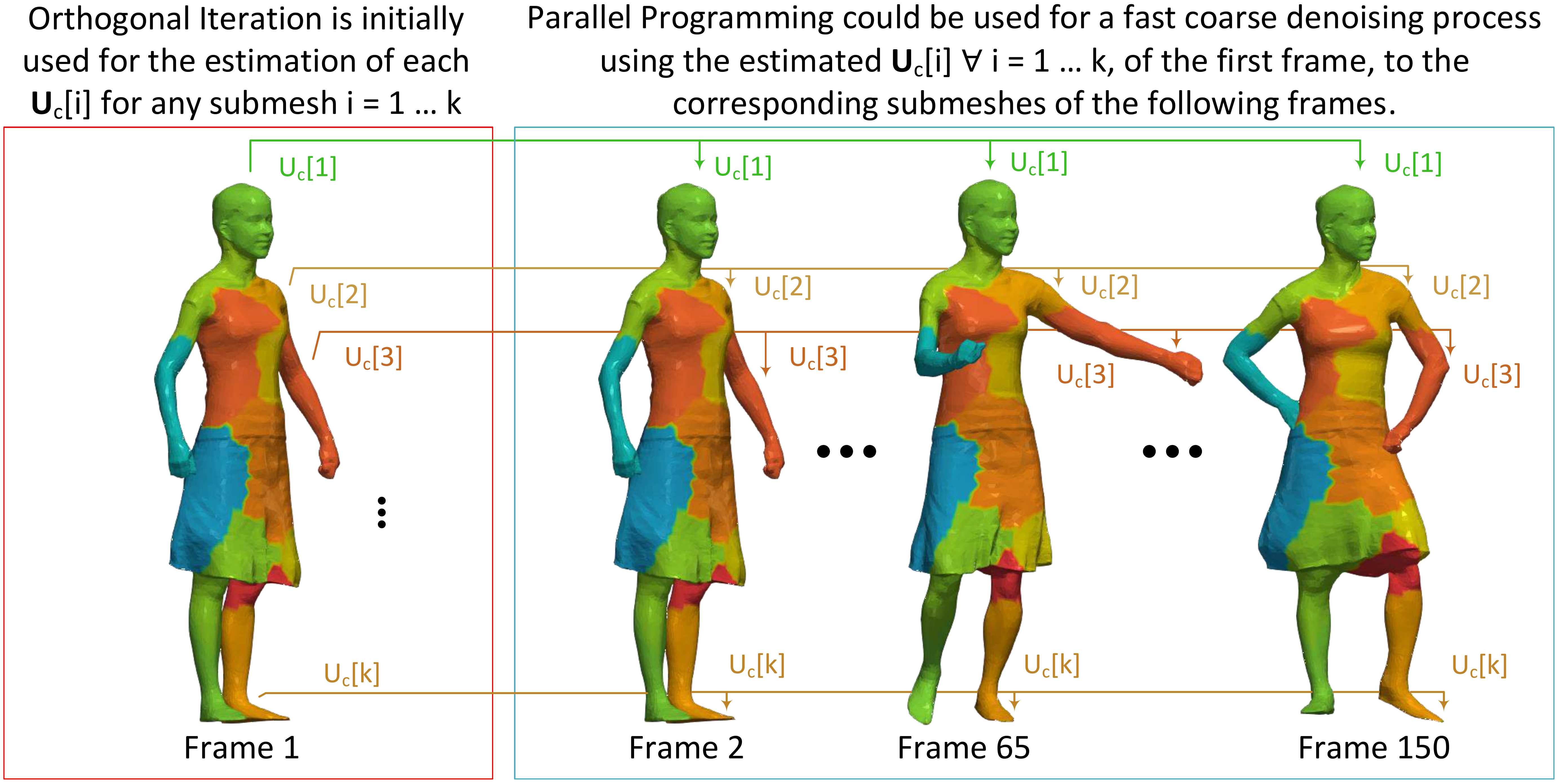}
	\mbox{}
	\caption{Parallel programming schema for high-performance coarse denoising of a 3D dynamic mesh.}\label{fig:parallel_programing_schema}
\end{figure}

Fig. \ref{fig:parallel_programing_schema} illustrates a schema representing the proposed coarse denoising of a dynamic mesh. The process starts by iteratively applying OI for the estimation of each $\mathbf{U}_c[i] \ \forall \ i = 1, ... , k$, as detailed described in Algorithm \ref{al:pipeline_AOI}. Then, parallel programming could be used for a fast coarse denoising process taking advantage of the already estimated matrices. In this case, the denoising process can run for all frames concurrently because no information of the previous frames is required  (except of the matrices $\mathbf{U}_c[i] \ \forall \ i = 1, ... , k$ which are estimated once during the OI process applied only to the first frame).

Additionally, adaptive compression of animated meshes could be used for real-time scenarios, as described in \cite{Lalos2017}.

\section{Performance Evaluation}

In the following section, we evaluate the presented framework in two different case studies: i) block based mesh compression and ii) block based mesh denoising, that effectively take advantage of the spectral coherence between different blocks utilizing OI. 

\begin{figure}
	\centering
	\includegraphics[width=0.82\linewidth]{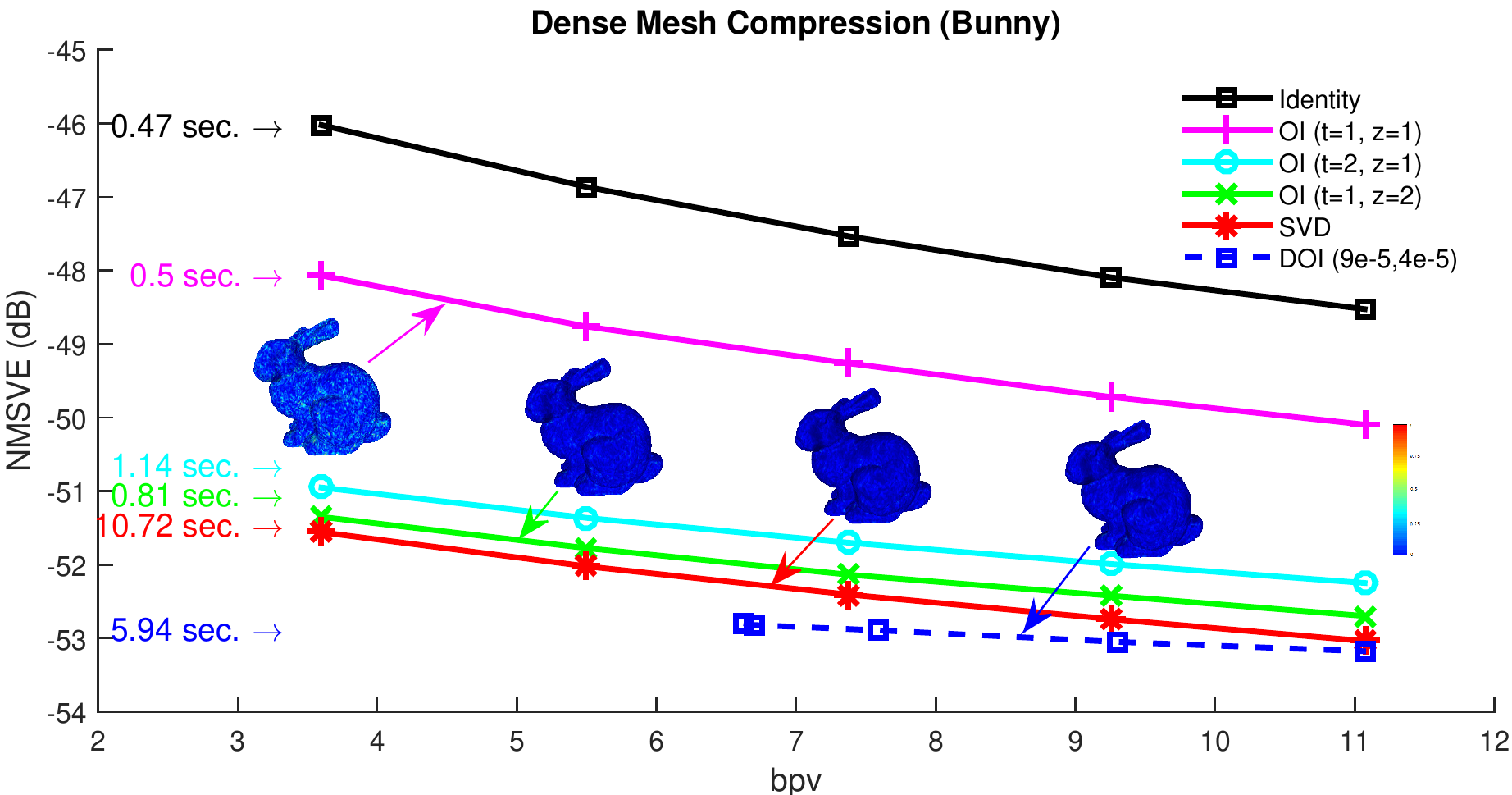}
	\caption{\label{fig:bunny_nmsven1}%
		Compression Study: NMSVE vs bpv for the Bunny model (34,817 vertices) was partitioned into 70 blocks with about 512 vertices per block.}
\end{figure}
\begin{figure}
	\begin{center}
		\includegraphics[width =0.83\linewidth]{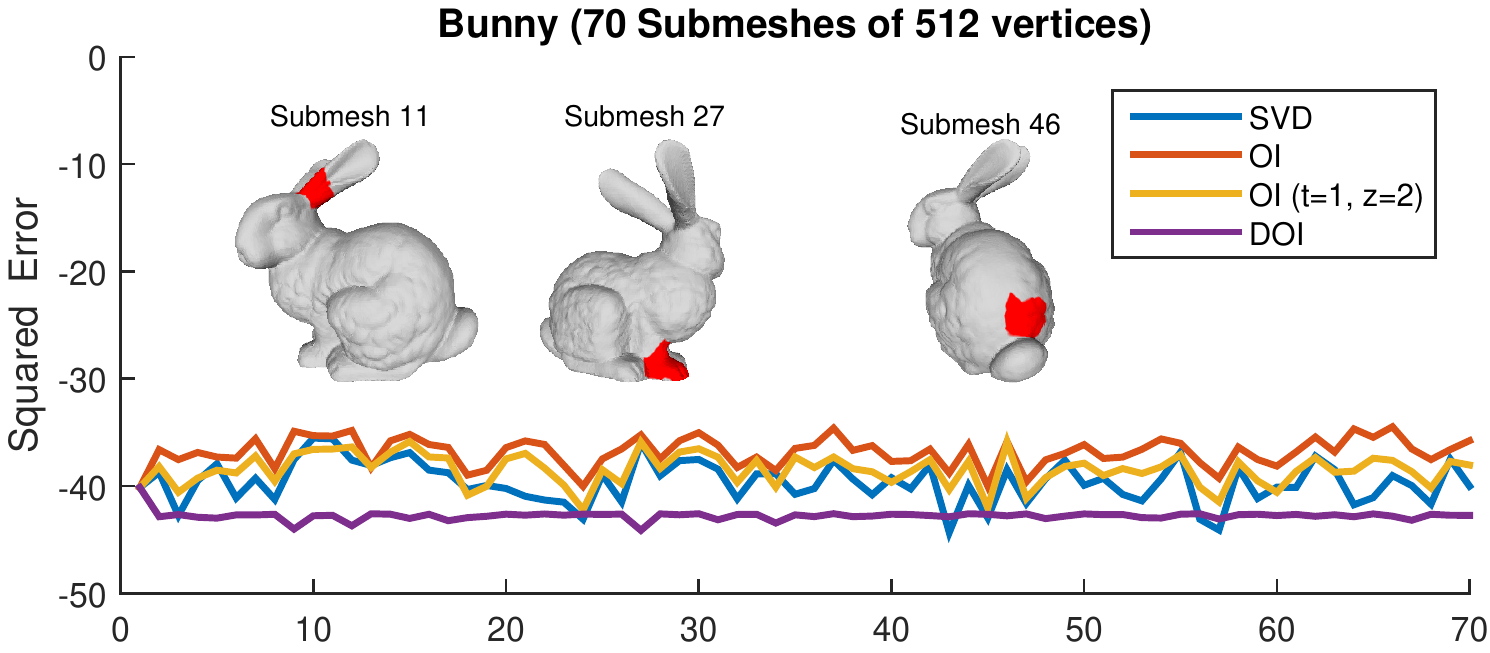}
		\includegraphics[width =0.83\linewidth]{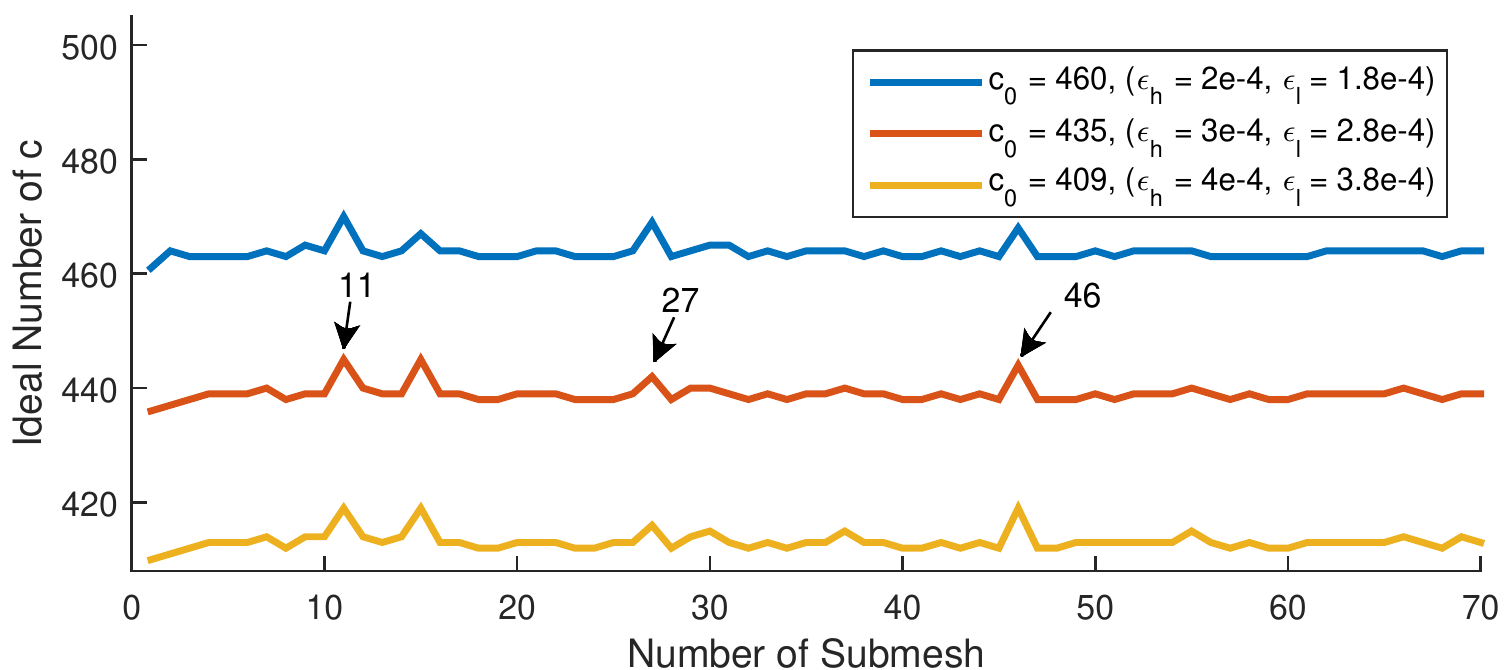}
		\caption{\textbf{[Up]} Squared Error per submesh for different approaches. \textbf{[Bottom]} Ideal value $c$ of subspace size per submesh.}
		\label{fig:fig2}
	\end{center}
\end{figure}

All simulations were performed on an Intel Core i7-4790 (3.6 GHz) processor with 8GB RAM. The compression efficiency of the geometry is measured in bits-per-vertex ($bpv= 3 \cdot q_c \cdot c \cdot k/n_d$) where $q_c$ are the bits used for uniformly quantizing the feature vectors ($q_c$ = 12 bits) and $c$ the total components kept from each submesh. This metric encapsulates the feature vectors for each processed block, ignoring the mesh connectivity which can be effectively compressed through any state-of-the-art connectivity encoder \cite{Rossignac1999}. To evaluate the reconstruction quality of our proposed method, it is necessary to capture the distortion between the original and the approximated frame. For this task, we chose the normalized mean square visual error (NMSVE) \cite{Karni2000} calculated as:
\begin{align}
\frac{1}{2n}\left(\left\|\mathbf{v} - \mathbf{\tilde{v}} \right\|_{l_2}+\left\|GL\left(\mathbf{v}\right) -GL\left(\mathbf{\tilde{v}} \right)\right\|_{l_2}\right)
\end{align}
where $GL\left(\mathbf{v}_i\right) = v_i - (\sum_{j\in N(i)}d^{-1}_{ij}v_j)/(\sum_{j\in N(i)}d^{-1}_{ij})$,
$\mathbf{v},\mathbf{\tilde{v}} \in \mathbb{R}^{3 n \times 1}$  represent vectors that contain the original and reconstructed vertices respectively, and $d_{ij}$ denotes the Euclidean distance between $i$ and $j$. 

\subsection{Compression Results}

The NMSVE vs bpv results are shown in Fig. \ref{fig:bunny_nmsven1} for the Bunny model. Note that the execution times shown next to each line encapsulate the respective time needed to construct $\mathbf{R}^z, z \geq1$, and to run the respective number of OI. By inspecting the figure, it can be easily concluded that the quality of the OI method performs almost the same as with SVD, especially when the number of iterations increases. At this point it should be noted that the benefits of our method are directly related to the size of each block. The theoretical complexities of the proposed schemes are in tandem with the measured times. More specifically, the OI approach for the Bunny mesh can be executed up to 20 times faster than the direct SVD approach. Although running more OI iterations yields a better NMSVE, converging towards the (optimal) SVD result, it comes at the cost of a linear increase in the decoding time. On the other hand, one iteration of $\mathbf{R}^2$ achieves lower visual error as executing two OI, in considerably less time. Moreover, DOI provides a stable reconstruction accuracy (see Fig. \ref{fig:fig2} showing the per submesh error) that can easily be adjusted by the defined thresholds. By inspecting also Fig. \ref{fig:fig2}, one can see that there is a coherence between submeshes since there are very few abrupt changes in the "ideal" value of subspace size that is required to satisfy a predefined reconstruction quality. 

However, this comes with a slight increase on the execution time (more OI) as well as a significant increase on the final compression rate (bpv) captured in Fig. \ref{fig:bunny_nmsven1} as a right shifting of the plot. The shifting is more obvious when the initial value of $c$ is small (more OI iterations are necessary for achieving the accuracy threshold).

\subsection{Denoising Results}

\begin{figure}[H]
	\centering
	\includegraphics[width=0.8\linewidth]{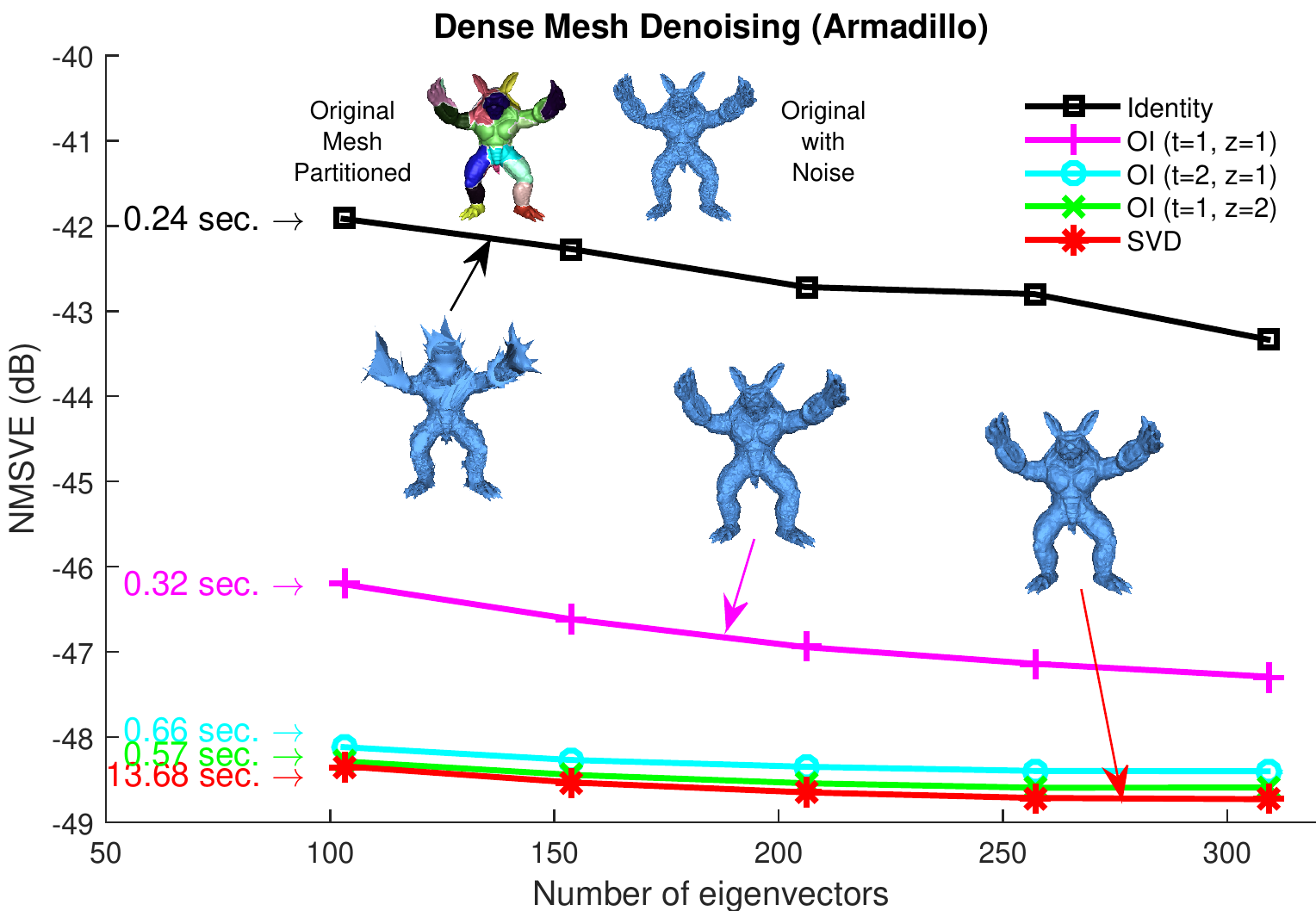}
	\caption{\label{fig:armadillo_nmsven1}%
		Coarse-to-fine Denoising study using the Armadillo model (20,002 vertices) was partitioned into 20 submeshes each comprising of around 990 vertices with zero-mean Gaussian noise $\mathcal{N}(0,0.2)$.}
\end{figure}

Similar conclusions are also drawn in a coarse-to-fine denoising setup where OI method are used as a pre-processing, "smoothing" step, before applying a conventional spectral bilateral filtering Fig. \ref{fig:armadillo_nmsven1}.

\begin{figure}[H]
	\begin{center}
		\includegraphics[width =0.80\linewidth]{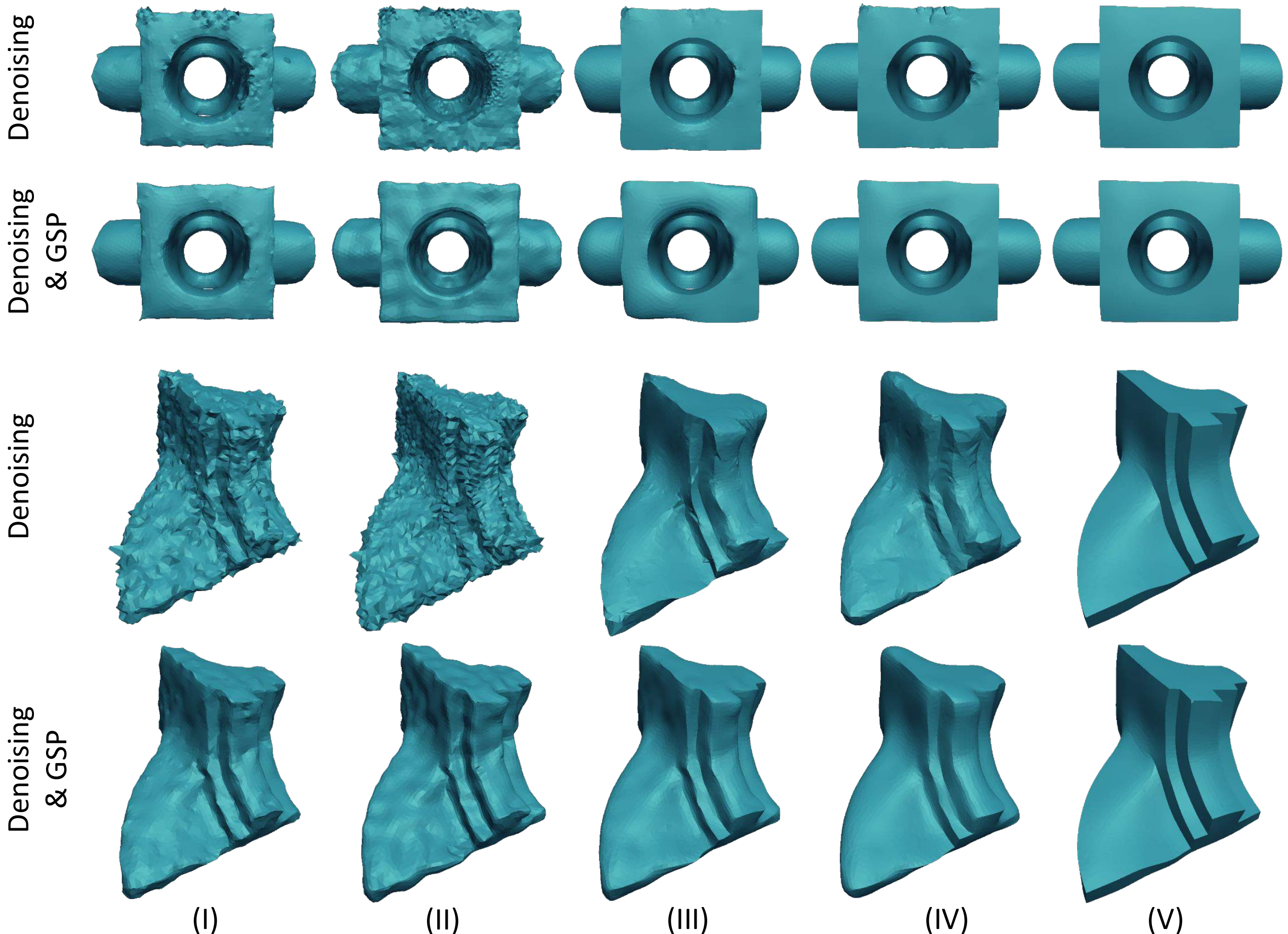}
		\caption{ Coarse denoising using Graph Spectral Processing (GSP) improves the efficiency of the following SoA approaches: (i) bilateral \cite{bilateral}, (ii) non iterative \cite{noniterative}, (iii) fast and effective \cite{fastandeffective}, (iv) bilateral normal [24], (v) guided normal filtering \cite{guidedfilter} (zero-mean Gaussian noise $\mathcal{N}(0,0.7)$).}
		\label{fig:fig3}
	\end{center}
\end{figure}

\begin{figure}[H]
	\begin{center}
		\includegraphics[width =0.90\linewidth]{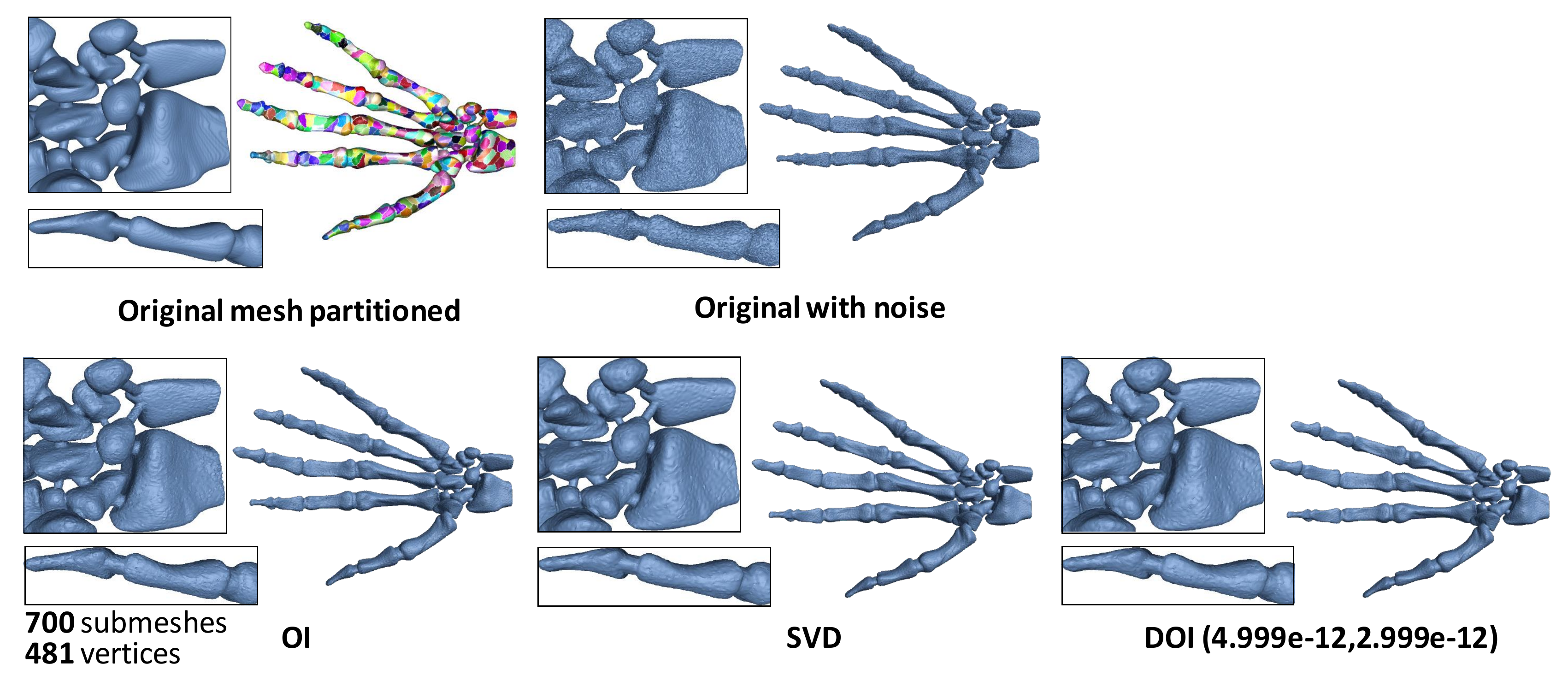}
		
		\caption{Graph Spectral processing of Hand (327,323 vertices) using $c=47$ eigenvectors, by applying the (a) OI method, (b) traditional SVD method, (c) Dynamic OI method.}
		\label{fig:fig4}
	\end{center}
\end{figure}

In Figs. \ref{fig:fig3} and \ref{fig:fig4}, it can be easily observed that the presented OI method can be employed by any other state-of-the-art (SoA) denoising method as a preprocessing step \cite{bilateralnormal},\cite{l0min},\cite{guidedfilter}, optimizing both its reconstruction quality and its computational complexity. The use of the coarse step significantly accelerates the convergence of the fine reducing the face/vertex update iterations required for achieving a specific reconstruction quality. The reconstruction benefits can be easily identified by inspecting Fig. \ref{fig:fig3} which presents denoising results of SoA methods (first row) and the corresponding results after using the OI approach as a preprocessing step (second row). 


\begin{figure}[H]
	\begin{center}
		\includegraphics[width =0.8\linewidth]{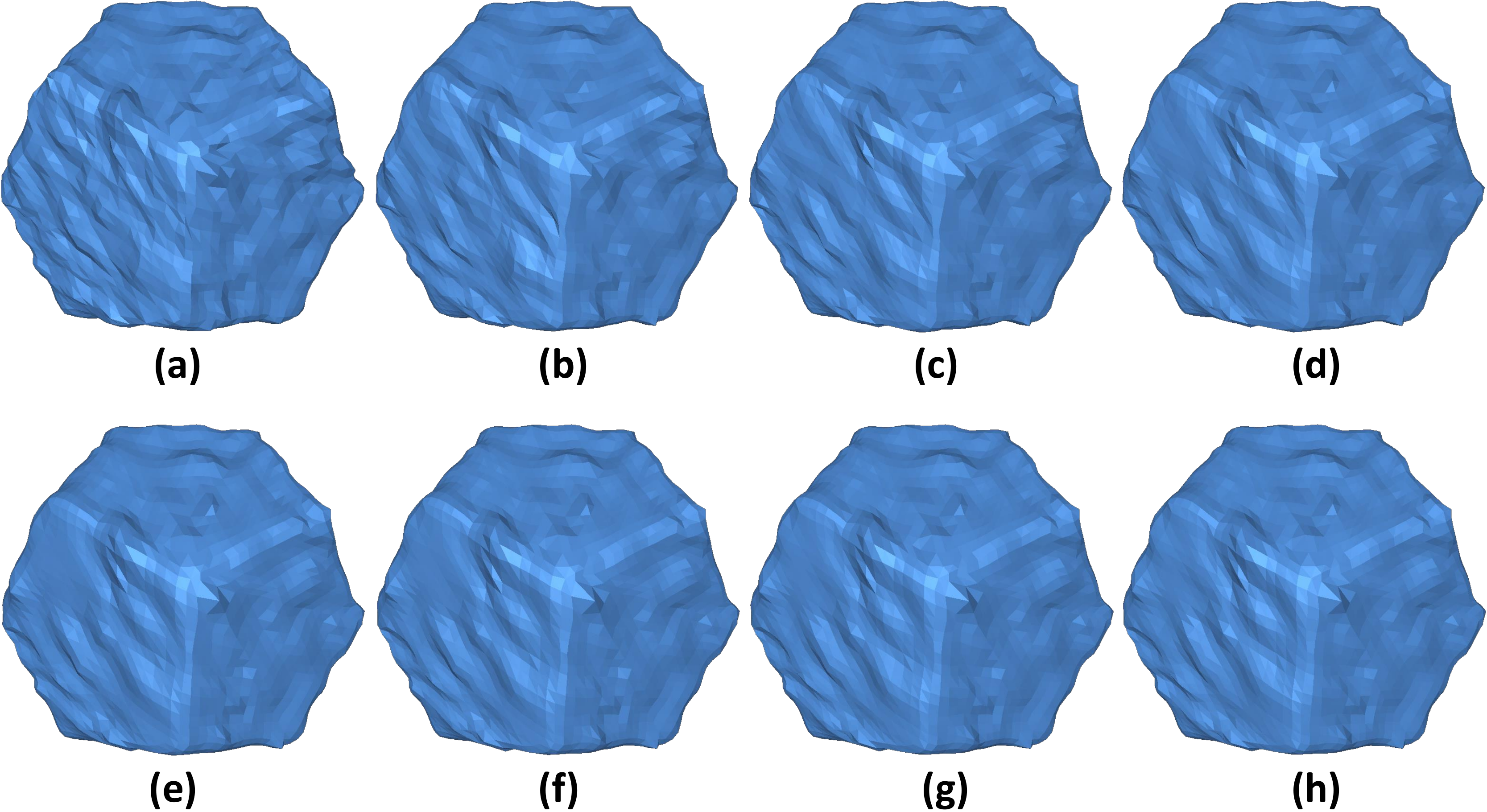}
		\caption{Coarse denoising results for different cases of $\mathbf{R}^z$, (a)
			$z=1$, (b) $z=2$, (c) $z=3$, (d) $z=4$, (e) $z=5$, (f) $z=6$, (g) $z=7$, (h) SVD.}
		\label{fig6}
	\end{center}
\end{figure}

Moreover, we also examined different combinations of $z$ (power of $\mathbf{R}$) and number of iterations in the denoising setup. Fig. \ref{fig6} shows the results of the coarse denoising step using OI for different values of $z$. Higher values, result in higher accuracies as compared to the direct application of SVD. While it should be noted that for $z > 4$ the results are identical with that achieved by applying the direct SVD. 

We evaluated the effects of executing several OI either on $\mathbf{R}^z$ or on $\mathbf{R}$ in CAD and scanned 3D models, using the NMSVE and  the angle difference between the normal of the ground truth face and the corresponding normal of the reconstructed face, averaged over all faces $\left(\theta\right)$. The differences between the SVD and OI, in terms of both reconstruction quality and execution time, are presented in Tables \ref{table:NMSVE_and_time} and \ref{table:Rpowertime}. It is clearly shown that the application of OI on $\mathbf{R}^z$ results in faster execution times than the application on $\mathbf{R}$. This result is attributed to the facts that: i) the execution of OI on $\mathbf{R}$ requires $z$ times more iterations to converge than the execution on $\mathbf{R}^z$ ii) while the evaluation of a matrix-matrix product (e.g., $\mathbf{R}\times \mathbf{R}$, e.t.c.) is much less computationally demanding than the orthonormalization step. These effects become more apparent in dense models. (see. Table \ref{table:NMSVE_and_time}).

\begin{table}[H]
	\caption{NMSVE and Execution Times} 
	\centering
	\begin{tabular}{|l|c|cc|ccc|} \hline
		\multicolumn{2}{|c|}{} &\multicolumn{2}{c|}{\rule{0pt}{12pt} \textbf{NMSVE (dB)}} & \multicolumn{3}{c|}{\rule{0pt}{12pt} \textbf{Time (sec.)}} \\ \hline 
		\rule{0pt}{12pt}	& \textbf{vertices}  & \textbf{SVD} &  \textbf{OI} & \textbf{SVD} & \textbf{OI} & \textbf{Speed-up}
		\\ \hline
		\multirow{4}{*}{\rotatebox{90}{\rule{0pt}{11pt} \textit{Armadillo}}} 
		\rule{0pt}{12pt}	& 10\% & -47.5668 & -47.4145 & 8.65 & 0.40 & \textcolor[rgb]{1,0,0}{22x} \\
		& 15\% & -47.6641 & -475263 & 8.80 & 0.53 & \textcolor[rgb]{1,0,0}{16x}\\
		& 20\% & -47.7428 & -47.6195 & 9.06 & 0.65 & \textcolor[rgb]{1,0,0}{14x}\\
		& 25\% & -47.8301 & -47.7097 & 9.16 & 0.80 & \textcolor[rgb]{1,0,0}{12x}\\
		\hline
		\multirow{4}{*}{\rotatebox{90}{\rule{0pt}{11pt} \textit{Hand}}} 
		\rule{0pt}{12pt}	& 10\% & -59.0354 & -58.8521 & 48.36 & 0.54 & \textcolor[rgb]{1,0,0}{89x}\\
		& 15\% & -58.872 & -58.7107 & 48.85 & 0.69 & \textcolor[rgb]{1,0,0}{70x}\\
		& 20\% & -58.6218 & -58.489 & 49.14 & 1.05 & \textcolor[rgb]{1,0,0}{46x}\\
		& 25\% & -58.3314 & -58.2215 & 49.53 & 1.31 & \textcolor[rgb]{1,0,0}{37x}\\
		\hline
	\end{tabular}
	\label{table:NMSVE_and_time}
\end{table}

Note that while both approaches, are based on a sequential update of the face normals and vertices, the GNF with coarse denoising, results to lower execution times (12x-89x). This reduction is attributed to the application of the coarse denoising step that filters out the high frequency components, accelerating the convergence speed of the necessary corrections/adjustments of the vertex positions.

\begin{table}
	\begin{center}
		\begin{tabular}{ | c | c | c ||   c | c | p{5cm} |}
			%\hline
			\cline{2-5}
			\multicolumn{1}{c|}{} & \multicolumn{2}{c||}  { \rule{0pt}{13pt} \textbf{Twelve}}  &  \multicolumn{2}{c|}{\textbf{Fandisk}} \\ \hline
			\rule{0pt}{13pt} & t & $\mathbf{\theta}$ & t & $\mathbf{\theta}$\\ \hline
			\rule{0pt}{13pt} \(\mathbf{R}^1\) & 0.031 & 11.57  & 0.077 & 16.36 \\ \hline
			\rule{0pt}{13pt} \(\mathbf{R}^2\) & 0.049 & 10.26 & 0.110 & 14.75 \\ \hline
			\rule{0pt}{13pt} \(\mathbf{R}^3\) & 0.099 & 13.97 & 0.170 & 14.54 \\ \hline
			\rule{0pt}{13pt} \(\mathbf{R}^4\) & 0.114 & 13.84 & 0.202 & 14.54 \\ \hline
			\rule{0pt}{13pt} \(\mathbf{R}^5\) & 0.136 & 13.7 & 0.242 & 14.44 \\ \hline
			\rule{0pt}{13pt} \(\mathbf{R}^6\) & 0.142 & 13.59 & 0.297 & 14.5 \\ \hline
			\rule{0pt}{13pt} \(\mathbf{R}^7\) & 0.157 & 13.57 & 0.313 & 14.52 \\ \hline
			\rule{0pt}{13pt} \(\mathbf{R}^8\) & 0.184 & 13.55 & 0.355 & 14.84 \\ \hline
			\rule{0pt}{13pt} \(\mathbf{R}^9\) & 0.201 & 13.53 & 0.407 & 15.6 \\ \hline
			\rule{0pt}{13pt} SVD & 0.901 & 9.83 & 1.953 & 14.56 \\ \hline
		\end{tabular}
		\vspace{0.3em}
		\caption {Execution time and face angle difference \(\theta\) for different cases of \(\mathbf{R}^z \) and SVD.}
		\label{table:Rpowertime}
	\end{center}
\end{table}

\section{Discussion and Conclusion}

In this paper, we introduced a fast and efficient way of performing spectral processing of 3D meshes ideally suited for real time applications. The proposed approach apply the problem of tracking graph Laplacian eigenspaces via orthogonal iterations, exploiting potential spectral coherences between adjacent parts. The thorough experimental study on a vast collection of 3D meshes that represent a wide range of CAD and scanned models showed that the subspace tracking approaches allow the robust estimation of dictionaries at significantly lower execution times compared to the direct SVD implementations. Despite the superiority of OI based approaches when compared to the direct SVD, the optimal subspace size should be carefully selected in order to simultaneously achieve the highest reconstruction quality and fastest compression times. 

\bibliographystyle{unsrt}
\bibliography{template}

\end{document}